\documentclass[12pt,onecolumn,letterpaper]{quantumarticle}

\pdfoutput=1
\usepackage{graphicx,import}
\usepackage[utf8]{inputenc}
\usepackage{layouts}
\usepackage{soul}
\include{colors}
\usepackage{tikz}
\usetikzlibrary{arrows}
\usetikzlibrary{positioning}
\usepackage[english]{babel}
\usepackage[T1]{fontenc}
\usepackage{amsmath}
\usepackage{hyperref,pdfpages}
\usepackage{hyperref,graphicx, amsmath,amssymb, amsthm, color, bm, bbm,dsfont,floatrow,enumerate,media9}
\usepackage[capitalise]{cleveref}
\usepackage[most]{tcolorbox}
\usepackage{graphicx}
\usepackage{caption}
\usepackage{array}
\usepackage{calc}
\newcolumntype{L}[1]{>{\raggedright\let\newline\\\arraybackslash\hspace{0pt}}m{#1}}
\newcolumntype{C}[1]{>{\centering\let\newline\\\arraybackslash\hspace{0pt}}m{#1}}
\newcolumntype{R}[1]{>{\raggedleft\let\newline\\\arraybackslash\hspace{0pt}}m{#1}}
\usepackage{siunitx}
\usepackage{cancel}

\usepackage{thmtools}
\usepackage{enumitem}
\usepackage{blindtext}
\declaretheoremstyle[
notefont=\bfseries, notebraces={}{},
bodyfont=\normalfont,
headformat=\NAME~\NUMBER:\NOTE
]{nopar}

\newtcbtheorem{boxthm}{Theorem}{
	enhanced,
	sharp corners,
	attach boxed title to top left={yshift=-\tcboxedtitleheight/2, xshift=3mm,yshifttext=-\tcboxedtitleheight/2},
	colback=white,
	colframe=red!20,
	coltitle=black,
	boxrule=0.7mm,
	fonttitle=\bfseries,
	boxed title style={
		sharp corners,
		size=small,
		colback=red!20,
		colframe=red!20,
	} 
}{thm}
\crefname{tcb@cnt@boxthm}{theorem}{theorems}

\newtcbtheorem{boxlem}{Lemma}{
	enhanced,
	sharp corners,
	attach boxed title to top left={yshift=-\tcboxedtitleheight/2, xshift=3mm,yshifttext=-\tcboxedtitleheight/2},
	colback=white,
	colframe=blue!20,
	coltitle=black,
	boxrule=0.7mm,
	fonttitle=\bfseries,
	boxed title style={
		sharp corners,
		size=small,
		colback=blue!20,
		colframe=blue!20,
	} 
}{lem}
\crefname{tcb@cnt@boxlem}{lemma}{lemmas}

\newtcbtheorem{boxprop}{Proposition}{
	enhanced,
	sharp corners,
	attach boxed title to top left={yshift=-\tcboxedtitleheight/2, xshift=3mm,yshifttext=-\tcboxedtitleheight/2},
	colback=white,
	colframe=blue!20,
	coltitle=black,
	boxrule=0.7mm,
	fonttitle=\bfseries,
	boxed title style={
		sharp corners,
		size=small,
		colback=blue!20,
		colframe=blue!20,
	} 
}{prop}
\crefname{tcb@cnt@boxprop}{proposition}{propositions}

\newtcbtheorem{boxdefn}{Definition}{
	enhanced,
	sharp corners,
	attach boxed title to top left={yshift=-\tcboxedtitleheight/2, xshift=3mm,yshifttext=-\tcboxedtitleheight/2},
	colback=white,
	colframe=green!55!black!15,
	coltitle=black,
	boxrule=0.7mm,
	fonttitle=\bfseries,
	boxed title style={
		sharp corners,
		size=small,
		colback=green!55!black!15,
		colframe=green!55!black!15,
	} 
}{defn}
\crefname{tcb@cnt@boxdefn}{definition}{definitions}

\newtcbtheorem{boxconj}{Conjecture}{
	enhanced,
	sharp corners,
	attach boxed title to top left={yshift=-\tcboxedtitleheight/2, xshift=3mm,yshifttext=-\tcboxedtitleheight/2},
	colback=white,
	colframe=yellow!20,
	coltitle=black,
	boxrule=0.7mm,
	fonttitle=\bfseries,
	boxed title style={
		sharp corners,
		size=small,
		colback=yellow!20,
		colframe=yellow!20,
	} 
}{conj}
\crefname{tcb@cnt@boxconj}{conjecture}{conjectures}

\newtcbtheorem{boxcoro}{Corollary}{
	enhanced,
	sharp corners,
	attach boxed title to top left={yshift=-\tcboxedtitleheight/2, xshift=3mm,yshifttext=-\tcboxedtitleheight/2},
	colback=white,
	colframe=red!20,
	coltitle=black,
	boxrule=0.7mm,
	fonttitle=\bfseries,
	boxed title style={
		sharp corners,
		size=small,
		colback=red!20,
		colframe=red!20,
	} 
}{coro}
\crefname{tcb@cnt@boxcoro}{corollary}{corollaries}

\newcounter{theo}

\crefname{theo}{theorem}{theorems}

\newcounter{lembx}

\crefname{lembx}{lemma}{lemmas}

\newcounter{defbx}

\crefname{defbx}{definition}{definitions}
\usepackage[framemethod=TikZ]{mdframed}
\usepackage[most]{tcolorbox}
\theoremstyle{plain}
\newtheorem{thm}{Theorem}

\theoremstyle{definition}
\newtheorem{defn}[thm]{Definition}
\newtheorem{rules}{Rules}

\crefname{thm}{theorem}{theorems}

\crefname{subroutine}{subroutine}{subroutines}
\crefname{protocol}{protocol}{protocols}

\definecolor{nblue}{rgb}{0.2,0.2,0.7}
\definecolor{ngreen}{rgb}{0.1,0.5,0.1}
\definecolor{nred}{rgb}{0.8,0.2,0.2}
\definecolor{nblack}{rgb}{0,0,0}

\DeclareMathOperator{\tr}{\mathrm{Tr}}

\usepackage{scalerel}
\usepackage{stackengine,wasysym}
\usepackage{pgf}

\newcommand{\hard}{{\rm hard}}
\newcommand{\echo}{{\rm echo}}

\providecommand{\ket}[1]{|#1 \rangle}

\newcommand{\mc}[1]{\mathcal{#1}}

\newcommand{\mbb}[1]{\mathbb{#1}}

\newcommand{\set}[1]{\mathbb{#1}} 

\newcommand{\avg}[1]{\left\langle {#1} \right \rangle}

\newcommand{\trans}[2]{t_{#1 \rightarrow #2}}

\definecolor{qbblue}{RGB}{38,151,208} 
\definecolor{darkblue}{RGB}{17, 42, 60} 
\definecolor{red}{RGB}{175, 49, 39} 

\definecolor{orange}{RGB}{217, 156, 55} 
\definecolor{qbgreen}{RGB}{66,184,99} 
\definecolor{palegreen}{RGB}{197, 184, 104} 
\definecolor{marigold}{RGB}{239,160,11}
\definecolor{ored}{RGB}{254,95,85}

\definecolor{yellow}{RGB}{250, 199, 100} 
\definecolor{brokenwhite}{RGB}{218, 192, 166} 
\definecolor{brokengrey}{rgb}{0.77, 0.76, 0.82} 

\newcommand{\hidden}[1]{}

\newcommand{\real}[0]{\set{R}\text{eal}}

\newcommand{\aoe}[1]{A\left(  #1 \right)}

\newcommand{\dd}[0]{{CarrPurcell1954,MG1958,Viola1998,Viola1999,VLK1999,Viola2003,KL2005,Uhrig2007,West2010,Yang2011}} 

\newcommand{\mitigation}[0]{{LiBenjamin2017,Temme2017,Dumitrescu2018,Endo2018,GT2020,He2020,Cai_2021a,Cai_2021b,Cai_2021d,Strickis2021,Huggins_2021,Koczor_2021a,Koczor_2021b,Obrien2021,Ferracin2022,HuoLi2022,cai_2022,Kurita2022}} 

\usepackage{tikz,qcircuit}
\usepackage{lipsum}
\usepackage{xcolor}
\def\pra{\ref@jnl{Phys.~Rev.~A}}        
\def\prb{\ref@jnl{Phys.~Rev.~B}}        
\def\prc{\ref@jnl{Phys.~Rev.~C}}        
\def\prd{\ref@jnl{Phys.~Rev.~D}}        
\def\pre{\ref@jnl{Phys.~Rev.~E}}        
\def\prl{\ref@jnl{Phys.~Rev.~Lett.}}    

\usepackage[numbers,sort&compress]{natbib}

\usepackage{mathtools}

\usepackage{subcaption}

\usepackage{color, colortbl, hyperref}

\usepackage[symbol]{footmisc}
\newcommand\correspondingauthor{\thanks{Corresponding author: padreher@ncsu.edu}}

\newcolumntype{C}[1]{>{\centering\arraybackslash}p{#1}}

\begin{document}

\title{Estimating Coherent Contributions to the Error Profile Using Cycle Error Reconstruction} 

\author{Arnaud Carignan-Dugas}
\affiliation{ Keysight Technologies Canada, Kanata, ON K2K 2W5, Canada }
\author{Shashank Kumar Ranu}
\affiliation{ Department of Electrical Engineering, Indian Institute of Technology Madras, Chennai 600036, India}
\affiliation{ Department of Physics, Indian Institute of Technology Madras, Chennai 600036, India}
\author{Patrick Dreher}
\affiliation{ Department of Electrical and Computer Engineering, North Carolina State University, Raleigh, NC  27695, USA}
\correspondingauthor{}

\maketitle

\begin{abstract}
{Mitigation and calibration schemes are central to maximize the computational reach of today's Noisy Intermediate Scale Quantum (NISQ) hardware, but these schemes are often specialized to exclusively address either coherent or decoherent error sources. Quantifying the two types of errors hence constitutes a desirable feature when it comes to benchmarking  error suppression tools. In this paper, we present a scalable and cycle-centric methodology for obtaining a detailed estimate of the coherent contribution to the error profile of a {hard} computing cycle. The protocol that we suggest is based on Cycle Error Reconstruction (CER), also known as K-body Noise Reconstruction (KNR).  This protocol is similar to Cycle Benchmarking (CB) in that it provides a cycle-centric diagnostic based on Pauli fidelity estimation \cite{erhard2019characterizing}. We introduce an additional hyper-parameter in CER by allowing the hard cycles to be folded multiple times before being subject to Pauli twirling. Performing CER for different values of our added hyper-parameter allows estimating the coherent error contributions through a generalization of the fidelity decay formula. We confirm the accuracy of our method through numerical simulations on a quantum simulator, and perform proof-of-concept experiments on three IBM chips, namely $\text{\texttt{ibmq\_guadalupe}}$, \texttt{ibmq\_manila}, and \texttt{ibmq\_montreal}. In all three experiments, we measure substantial coherent errors biased in $Z$.}
\end{abstract}

\section{Introduction}

Today's engineering advances in quantum computing hardware architectures allow vendors to produce processors with a considerable number of qubits.  Despite these impressive advances, the current technology does not yet enable large-scale fault-tolerant quantum computing (FTQC) to be implemented. As a result, computations today and into the intermediate future will be done on Noisy Intermediate-Scale Quantum (NISQ) \cite{preskill2018quantum} platforms.  Noise on these devices can result in both decoherent and coherent errors. Suppressing noise on these hardware platforms and extending the coherence times will rely on improvements in error mitigation protocols and error suppression methods. 

Decoherent errors are created by the unwanted interactions between the qubits and their external environment as well as by stochastic events. Typical decoherent processes include amplitude-damping, phase-damping, and depolarization. Decoherent errors are often partially suppressed by dynamical decoupling (DD) techniques \cite{\dd}. Moreover, many error mitigation schemes rely on the propagation laws of decoherent errors to suppress their effect on computational outcomes \cite{\mitigation}.

Coherent errors are the result of unitary processes occurring in the closed system formed by the qubits' Hilbert space. They arise from several sources such as mis-calibrated control parameters, external fields, and crosstalk (e.g. undesired internal fields). They are usually suppressed through advanced calibration and pulse compensation methods \cite{bultrini2020simple,Winick_2021, Ahsan_2022,CER_2023}. Alternatively, coherent errors can be effectively transformed into decoherent errors via Randomized Compiling (RC) \cite{wallman2014randomized,Winick-2022}, and then mitigated with tools aimed for decoherent errors \cite{Kurita2022}.

Because the two types of errors have different natures, there are specific error suppression tools {aimed} at suppressing only coherent errors, while others impact decoherent errors either exclusively, or more efficiently.  Given the fundamentally different nature of coherent and decoherent errors and the likelihood that FTQC will not be available in the intermediate future, it will be important to develop fast and accurate methods to independently measure their impact on quantum computing outcomes. Developing such capabilities can provide an essential means to test and compare the effect of different error suppression and calibration tools. 

There exist many different frameworks and protocols for characterizing quantum computing processing errors. One approach focuses on characterizing specific infinitesimal generators of motion from a Hamiltonian or Lindbladian viewpoint. While crucial, such approaches don't immediately connect with the diagnostics of integrated evolutions such as quantum gates. Let's define a quantum instruction as a set of one or few qudit labels, together with a linear operation to apply over these. The operation can be a gate, some measurement operation, or some state preparation. A cycle can be defined as a set of instructions targeting disjoint sets of qubits, together with a schedule. For example, a cycle could consist of a round of simultaneous CNOT gates over the processor. Due to various physical effects such as crosstalk, parallel instructions often substantially influence each other. This observation means that characterizing instructions in isolation can lead to misguided diagnostics.  Indeed, a cycle error channel depends on a non-obvious function of every instruction that is being applied, as well as on the precise relative timing of those instructions.

As such, to better contextualize error profiles, we focus on a cycle-centric approach to process characterization. Given the considerable number of parallelizable instructions as well as the complexity of their joint error profile, a cycle-centric approach is particularly relevant in the NISQ-era and beyond. A well-known cycle-centric error diagnostic scheme is known as Gate Set Tomography (GST) \cite{Merkel2013,Blume-Kohout2013,Blume_Kohout_2017,Greenbaum2015, Nielsen_2021,Brieger2021}. GST provides detailed information about both coherent and decoherent errors, but the diagnostic information addresses non-randomized cycles. 

In the current work, we instead focus on characterizing effective dressed cycles tied to randomly compiled circuits.  The idea is to provide detailed diagnostic information that closely connects with the effective performance of circuits run under RC.  One convenient aspect of effective dressed cycles is that they can be characterized via Pauli infidelity estimation methods \cite{FW2019,HWFW2019}. 
In particular, the Cycle Benchmarking (CB) \cite{erhard2019characterizing} protocol features a valuable framework to characterize effective dressed cycles with great precision. As such, more advanced error diagnostic tools, such as Cycle Error Reconstruction (CER) \cite{CER_2023,cer_aps}, repurposed the structure of CB circuits to gather detailed error profiles attached to effective dressed cycles. Note that CER is also known as $k$-body Noise Reconstruction (KNR) \cite{knrpatent2018}.

{
However, neither CB nor CER are initially designed to provide a budget for coherent and decoherent contributions to the measured error rates. As such, we introduce a new CB-structured method for separately measuring both the decoherent and main coherent cycle errors based on CER data. To distinguish between decoherent and coherent errors, we rely on their different propagation principles (linear vs quadratic). In the current work, we focus on characterizing effective dressed cycles tied to randomly compiled circuits. 
}

{
The idea is to provide detailed diagnostic information that closely connects with the effective performance of circuits run under RC.  This technique can be as fine as desired, depending on the number of marginal error probabilities that are included. One important purpose of making the distinction between coherent and decoherent errors (other than understanding the error budget in a more complete manner) is that this can lead to calibration strategies such as knowledge-based dynamical decoupling or compiled compensations of coherent errors into future cycles. 
}

{
We acknowledge that both GST and our extended CER do provide a measurement of the noise. However, CER considers effective dressed cycles which are slightly different objects than deterministic cycles characterized in the GST method. Of course, one could deduce the error maps of effective dressed cycles by combining various GST-obtained error profiles, but the advantage of the extended CER is that it leverages the tailored properties of the noise under RC to accelerate the characterization procedure.
}

In this paper we layout our work as follows. In \cref{sec:theory-section}, we summarize the main principles describing the propagation of coherent and decoherent errors in repeated error channels. In \cref{sec:CER protocol}, we outline the background material necessary to understand a CB-structured error characterization scheme known as CER \cite{CER_2023,cer_aps}, and we introduce our protocol extension to CER. We finish the section with a generalized decay model that connects the output data of the scheme to the effective coherent contributions to the error profile {of the hard cycle}. In \cref{sec:Magnon-example}, we further expand on the fitting model related to our protocol. This section also provides numerical evidence that the error profile obtained from our suggested experiment matches the underlying error model. In \cref{data analysis}, we apply our extension to the CER protocol on three separate IBM quantum computing hardware platforms and present the central results of this research.  As a proof of concept, we measure the effective coherent and decoherent error rates marginalized on an ancilla during an entangling cycle. Fidelities of the errors for various sequence lengths are calculated and compared to the results obtained from running the circuit applying a proposed polynomial global fitting function. We show through our implementations that the coherent error manifests primarily as Pauli $Z$ error on IBM platforms due to $ZZ$ static coupling in transmon qubits.  In \cref{sec:Summary}, we summarize 
our results and expand on their relevance regarding the benchmarking and designing of error suppression tools for gate-based quantum computing architectures.  In addition, there are several supplemental sections that expand the discussion in the main sections of the paper. 

\section{Theory - Coherent and decoherent error propagation}
\label{sec:theory-section}

Coherent and decoherent quantum errors in a quantum computing circuit propagate in intrinsically different ways.  In this section, we explicitly demonstrate the error propagation formula for the simple case of an $x$-fold error channel $\mc{E}^x$.  

 Recent work has also analyzed the error propagation in quantum computers \cite{yu2022analysis}. They have shown that the decoherent error increases linearly with circuit depth but plateaus after some threshold circuit depth. Our work takes such an analysis into coherent error domain, and shows, both theoretically as well as through implementations on IBM quantum processor, that repeated coherent error propagates quadratically to the first order.

A quantum state can be described by a density matrix $\rho$, which is a positive semi-definite, unit-trace matrix. This allows representing pure states (if and only if the rank of $\rho$ is $1$) and a probabilistic mixture thereof. The evolution of a quantum mechanical state $\rho$ is often described by the von Neumann equation, 
\begin{align}
    \frac{d}{dt} \rho = -i[H,\rho]~,
\label{eq:Schrodinger}
\end{align}
where $H$ is a Hermitian matrix referred to as the Hamiltonian. However, the von Neumann equation only applies to the deterministic evolution of a closed system. 

For open quantum systems, a more general evolution formula is given by the Lindblad master equation \cite{Lindblad1976}:
\begin{align}
    \frac{d}{dt} \rho = & \Lambda[\rho] \notag \\
    :=& -i[H,\rho] + \sum_j \left( L_j \rho L_j^\dagger - \frac{1}{2} (L_j^\dagger L_j\rho  + \rho L_j^\dagger L_j)  \right)~,
\label{eq:Lindblad}
\end{align}
where $L_i$ are traceless matrices referred to as Lindblad operators or Lindblad jumps, and where the linear operator $\Lambda$ is referred to as the Lindbladian. A Lindlbadian is one of the general forms of Markovian and time-homogeneous master equations describing the general non-unitary evolution of the density matrix $\rho$.

An error channel $\mc E$ can be represented as a linear operator acting on density matrices.  In this work, we consider error channels $\mc E$ which are close to the identity and that are the result of a finite-time evolution of a time-independent Lindbladian: 
\begin{align}\label{eq:integral}
    \mc E [\rho] = \int_{t=0}^{t= \Delta t} \Lambda[\rho] dt~.
\end{align}
For what follows, we pick $\Delta t =1$ without loss of generality, as it simply fixes the scales of the matrices $H$ and $\{L_i\}_i$ appearing in the Lindbladian. We emphasize that although the $d \times d$ matrices ($d=2^{n}$ where $n$ is the number of qubits) $H$ and $L_j$ appearing in \cref{eq:Lindblad} do not depend on time, we don't have to assume that the error channel $\mc E$ is the result of a time-independent error process. The error process itself could involve time-dependent features, but could be equivalent to an integral over a time-independent evolution.  Indeed, if the error process is weak (which implies that the generators of motions can be time-averaged) then after the time-dependent equation is integrated, the solution can be either exactly expressed or well approximated as the integral of a time-independent process. 

Starting from ~\cref{eq:integral} we can construct a columned-vectorized picture (i.e. the $d \times d$ density matrix $\rho$ becomes a $d^2 \times 1$ vector $\rm{col}({\rho} )$ obtained by stacking the columns of $\rho$) and re-write the equation as
\begin{align} \label{eq:col_master}
    \frac{d}{dt} \rm{col}({\rho})= \Lambda~ \rm{col}({\rho})~,
\end{align}
where the $d^2 \times d^2$ matrix

\begin{align}
    \Lambda := -i (\mbb{I} \otimes H-H^T\otimes \mbb{I} ) +
    \sum_j\left[L_j^* \otimes L_j -\frac{1}{2}\left( \mbb{I} \otimes L_j^\dagger L_j + L_j^T L_j^* \otimes \mbb {I}\right)\right]~
\label{eq:lambda}
\end{align}

is the matrix form of the Lindbladian. The solution of \cref{eq:col_master}
is given by \cref{eq:density}.

{
\begin{equation}
{\rm{col}}({\rho(t)}) = e^{\Lambda (t-t_{0})} {\rm{col}}(\rho(t_{0}))
\label{eq:density}
\end{equation}
}
Therefore our error channel $\mc E$ can be expressed as
\begin{align}
    \mc E = e^{\Lambda}~.
\end{align}
In this context we are viewing $\Lambda$ as the generator of the process matrix $\mc E$. Viewing $\Lambda$ from this perspective, if all of the terms in $\Lambda$ can be calculated, then effectively one knows the error process matrix $\mc E$.

To properly {discuss} the strength of specific terms appearing in the Lindbladian, let's fix an expansion basis and express the effective Hamiltonian $H$ and Lindblad operators $L_j$ as linear combinations Pauli operators:
\begin{subequations}
    \begin{align}
 H &=: \sum_{P \in \set{P}_n} h_P P~,\label{eq:H} \\
 L_j&=: \sum_{P \in \set{P}_n} \ell_{j,P} P~,
\label{eq:L}
\end{align}
\end{subequations}
{where $\set{P}_n$ denotes the $n$-qubit Pauli basis defined by the Kronecker products of the Pauli matrices $I, X, Y, Z$}. The coefficients obey $h_P\in \set{R}$, $\ell_{j,P}\in \set{C}$. In practice, the squared magnitudes $|h_P|^2$ and $|\ell_{j,P}|^2$ are upper-bounded by the 2-qubit error rates because they are ultimately tied to geometric near-local interactions.

In this work, we only consider errors induced by near-local physical interactions. To provide a notion of locality, we have to consider the geometric features of a given device. We will simply consider an interaction graph, where qubits can undergo controlled entangling operations if the vertices are connected. Given a connectivity graph, we constrain Lindblad and Hamiltonian operators to be at most geometrically $k$-local for some small integer $k$, which means that the operators can act on connected subgraphs of at most $k$ qubits (note that the operators can act trivially on some of the qubits, allowing for gaps). The accuracy of the upcoming \cref{eq:fidel-repeated-channel} fails exponentially fast in the size of the locality constraint $k$.  However, given some array-like topology, if we assume that errors stem from 1 or 2-body interactions in some rotating frame, then an N-qubit subsystem ($N \le n)$ should be affected by $O(N)$ terms in the Lindbladian. As a result (see \cref{sec:appendix A}), the action of the error channel $e^{\Lambda}$ on a limited qubit support can be well approximated through an early-truncated Taylor expansion. 

To  be more specific,  first consider \emph{Pauli fidelities}, defined for a generic channel $\mc E$ acting on a $n$-qubit system as
\begin{align}
    f_P(\mc E):= \frac{\tr P \mc E[P]}{2^n}~.
\label{eq:pauli_fid}
\end{align}

We denote the infidelity as $\delta f_P(\mc E) = 1-f_P(\mc E)$. Given a Lindbladian $\Lambda$ such that $\mc E = e^\Lambda$, let's further define the coherent and decoherent contributions to the infidelities respectively as:

\begin{subequations}\label{eq:infidelities}
    \begin{align}
     \delta f_P^{\rm coh.} := 2\sum_{\substack{Q\\ QP=-PQ}}|h_{Q}|^2 \\
     \delta f_P^{\rm decoh.} := 2\sum_{\substack{j, Q\\ QP=-PQ}} |\ell_{j,Q}|^2
    \end{align}
\end{subequations}

In this work, we approach the propagation of errors by studying the Pauli fidelities associated with the repeated channel $\mc{E}^x =e^{x\Lambda}$:
\begin{align} 
       f_P(\mc E^x) 
       & = 1-  \delta f_P^{\rm coh.}x^2 - \delta f_P^{\rm decoh.} x \notag \\
       &+O\left(\delta f^{\rm decoh}_P\left(\delta f^{\rm decoh}_P +\sqrt{\delta f^{\rm coh}_P}\right)x^2\right)+O(x^3)~
\label{eq:fidel-repeated-channel}
\end{align} 
Notice that both sums above are performed over the Paulis which anti-commute with $P$, and the second sum is further performed over the Lindblad jumps' indices $j$.
In \cref{sec:appendix A}, we provide a proof of \cref{eq:fidel-repeated-channel}, and expand on the quadratic term in $x$. 

If we Pauli-twirl the error $\mc E$ -- and denote it as $\mc E^{\rm RC}$ to refer to randomized compiling -- we could sensibly characterize it in terms of Pauli error probabilities, since the twirled error channel would become a probabilistic sum over Pauli errors:

\begin{align}
    \mc E^{\rm RC}[\rho]  = \sum_{P \in \set P_n} e_P P \rho P^\dagger~,
\end{align}

where $\set P_n$ is the $n$-qubit Pauli {basis}, and where $\{e_P\}_P$ is a probability distribution over Pauli errors $P$. We can translate between Pauli error probabilities $e_P$ and Pauli fidelities $f_P$ by using the Walsh-Hadamard transform (see \cite{NC2010} as well as \cref{wh_transform}). We can define the coherent and decoherent contributions to the error probability $e_P$ respectively as:

\begin{subequations}
    \begin{align}\label{eq:error_rates}
        e_P^{\rm decoh.} &:= \sum_j|\ell_{j,P}|^2 \\
        e_P^{\rm coh.} &:= |h_P|^2 
    \end{align}
\end{subequations}

Notice that by substituting the above in \cref{eq:infidelities}, we get
\begin{subequations}\label{eq:infidelities_as_error_rates}
    \begin{align}
     \delta f_P^{\rm coh.} &= 2\sum_{\substack{Q\\ QP=-PQ}} e_Q^{\rm coh.} \\
     \delta f_P^{\rm decoh.} &= 2\sum_{\substack{ Q\\ QP=-PQ}} e_Q^{\rm decoh.}
    \end{align}
\end{subequations}
Just as in \cref{eq:fidel-repeated-channel}, error probabilities corresponding to 
a Pauli-twirled repeated channel $(\mc E^x)^{\rm RC}$ closely follow a quadratic equation:
  {\begin{align} 
       e_P(\mc E^x) &\simeq x^2 ~e_P^{\rm coh.} + x~ e_P^{\rm decoh.}   \label{eq:quad-linear-errors}
   \end{align}}
We elaborate on \cref{eq:quad-linear-errors} in 
\cref{sec:appendix B}. To a good approximation, coherent errors first induce a purely quadratic error growth as opposed to the purely linear initial growth characteristic of decoherent errors. This difference in propagation speed is the key to differentiating the two error sources.

\section{Learning the Error Profile Attached to a Cycle}
\label{sec:CER protocol}

In this section, we provide the background material necessary to understand a CB-structured error characterization scheme known as Cycle Error Reconstruction (CER), which is also known as $k$-body Noise Reconstruction (KNR). Once the basics of CER are covered, we shall introduce our extension to the protocol which complements the standard CER diagnostics with a measure of the coherent contribution to the error profile {of the hard cycle}.

\subsection{Cycle Error Reconstruction (CER) Error Diagnostic Protocol}\label{subsec:cer}

In quantum computing, there are typically hard and easy cycles, which usually consist of parallel entangling gates and parallel single-qubit gates respectively. Hard cycles dictate the circuit performance compared to easy cycles. A dressed cycle refers to a composite cycle comprising a hard cycle followed (or preceded) by an easy cycle.

CER specifically provides error profiles attached to cycles relevant to circuits performed under Randomized Compiling (RC), a well-known error suppression technique \cite{wallman2016noise,PhysRevX.11.041039,ville2021leveraging,gu2022noise,winick2022concepts}. To understand RC, consider a noisy application circuit composed of $m$ sequential dressed cycles $C_{i}$ interleaved with cycle-dependent error channels $\mc E_i$,
\begin{align}
    \tilde{\mc C} = \mc E_m C_m \cdots \mc E_1 C_1~.
\end{align}
RC replaces the above application circuit $\tilde{\mc C}$ meant to be run $N_{\rm shots}$ {times with $n_{\rm rand}$} randomly sampled equivalent circuits (over a structured distribution of circuits), each to be run $N_{\rm shots}/n_{\rm rand}$ times. The effect of RC is to effectively replace each noisy dressed cycle $\mc E_i C_i$ with what is 
referred to as an \emph{effective dressed cycle} $\mc E_i^{\rm RC} C_i$, which is similar to $\mc E_i C_i$. The difference here is that the generic error channel $\mc E_i$ is replaced with a Pauli stochastic error channel $\mc E_{i}^{\rm RC}$ of the form 

\begin{align}
    \mc E_{i}^{\rm RC} [\rho]  = \sum_{P \in \set P_n} e_{P,i} P \rho P^\dagger~,
\end{align}
where $\set P_n$ is the $n$-qubit Pauli {basis}, and where $\{e_{P,i}\}_P$ is a probability distribution over Pauli errors $P$ corresponding with the {$i$\rm{th}} cycle. The error distribution naturally depends on the cycle $C_i$. 

CER specifically learns information about the distribution $\{e_{P,i}\}_P$. To avoid scalability limitations, CER doesn't usually try to learn every probability value $e_{P,i}$, because there are $4^n$ of them. Instead, it learns marginal error probabilities over constrained unions of instruction labels. That is, if we let $\set A$ be a union of instruction labels (e.g. $\{1\} \cup \{2,3\}$), and let $\set A^c$ be its complement over the architecture (e.g. $\{4, \cdots, n\}$), then the marginal probability of the error $P$ (e.g. $XXZ$) over $\set A$ is defined as

\begin{align}
    \mu_{\set A}(P):= \sum_{Q} e_{\text{$P$ over $\set A$ and $Q$ over $\set A^c$}}~.
\end{align}

CER is tomographically complete because learning all marginals is equivalent to learning the probability distribution $\{e_{P,i}\}_P$. However, due to locality constraints in realistic error models, ${\rm poly} (n)$ marginals are often sufficient to accurately reconstruct the error profile of a given cycle.
For instance, in a planar architecture, if we look at the marginal error probabilities for all pairs of gates in a given cycle that are occurring on two directly connected sets of qubits, we would only need to gather $O(C \cdot n)$ error probabilities, where $C$ denotes the average number of connections per qubit (in a rectangular lattice, $C = 4$). If we want to further collect next-to-nearest neighbor error correlations, we would need $O(C^2\cdot n)$ error estimates. 
Keeping the same two examples but for a fully connected architecture, the costs would instead be $O(n^2)$ and $O(n^3)$ respectively. A more thorough discussion regarding marginal probabilities and reduced error models in CER can be found in \cite{CER_2023}.

The marginal error probabilities associated with the stochastic channel $\mc E_{i}^{\rm RC}$ are retrieved from the \emph{Pauli fidelities}, defined in \cref{eq:pauli_fid}. That is, given an effective cycle of interest $\mc E^{\rm RC} C$, the CER protocol involves the estimation of a set of various Pauli fidelities, {$\{f_P(\mc E^{\rm RC})\}_P$}, and converts the cycle of interest into a marginal error distribution. The Pauli fidelities are themselves extracted from circuit-level observables. 

To be more explicit, let's specify CER circuits with three parameters, namely a set of commuting Paulis, $\set S$, which is dictated by a state preparation and measurement (SPAM) strategy, a number of dressed cycle repetitions ``$m$'', and a string ``$s$'' sampled from a random variable $X$, which contains the information about the RC-randomized dressings. The role of the three parameters can be visualized in \cref{fig:circuit_representation} (set $x=1$), where the random string $s$ dictates a choice of easy cycles. Each sampled CER circuit $\tilde{\mc C} (\set S, m, s)$ yields a set of fidelity-like numbers $\{f^{\rm circuit}_P(m , s)\}_{P \in \set S}$ associated with the set $\set S$ of commuting Pauli operators determined by the SPAM strategy. The expectation of $f^{\rm circuit}_P(m , s)$ over the strings $s$ in $X$ obeys a decay formula of the form:
\begin{align}\label{eq:CER_decay}
 \mbb E_{s} f^{\rm circuit}_P(m , s) = A_P \left(f_P(\mc E^{\rm RC}) \right)^m~,
\end{align}
where the constant $A_P$ depends on SPAM errors, and where $\mc E^{\rm RC}$ is the Pauli stochastic error channel associated with the effective dressed cycle of interest. In practice, the LHS of \cref{eq:CER_decay} is replaced by a sample mean, and the Pauli fidelities $f_P(\mc E^{\rm RC})$ are obtained by gathering sample mean estimators for various values of dressed cycle repetitions $m$. {A thorough description of the CER protocol together with pseudo-code can be found in \cite{CER_2023}.}

\subsection{An extension to CER to learn coherent contributions to errors}
\label{subsec:CER-error}

CER is meant to provide a diagnostic of effective RC cycles. Let the implementation of a hard cycle be expressed as the product $ \mc E_\hard \circ C_\hard$, where $\mc E_\hard$ is the cycle error channel attached to the ideal cycle implementation $C_\hard$. Similarly, let $n$-qubit Pauli easy cycles $Q \in \set P_n$ be expressed as the product $Q \circ \mc E_Q $ (error channels can be placed on the left or right of the ideal implementation without loss of generality, but may differ depending on which side is chosen).  To a high approximation, the fidelities given by CER correspond to \cite{CER_2023}:
\begin{align}\label{eq:cer_fid}
   f_P^{\rm CER} = f_P\left(  \avg{\mc E_{Q}}_{Q \in \set P_n} \circ \mc E_\hard \right)~,
\end{align}
and it corresponds to the Pauli fidelities of the effective dressed cycle $\mc E^{RC} \circ C_\hard$ to the ideal cycle $C_\hard$.  Notice that those fidelities don't provide a quantitative budgeting of the coherent and decoherent contributions to the infidelity.

\begin{figure}[h!]
\centering
\def\svgwidth{\columnwidth}
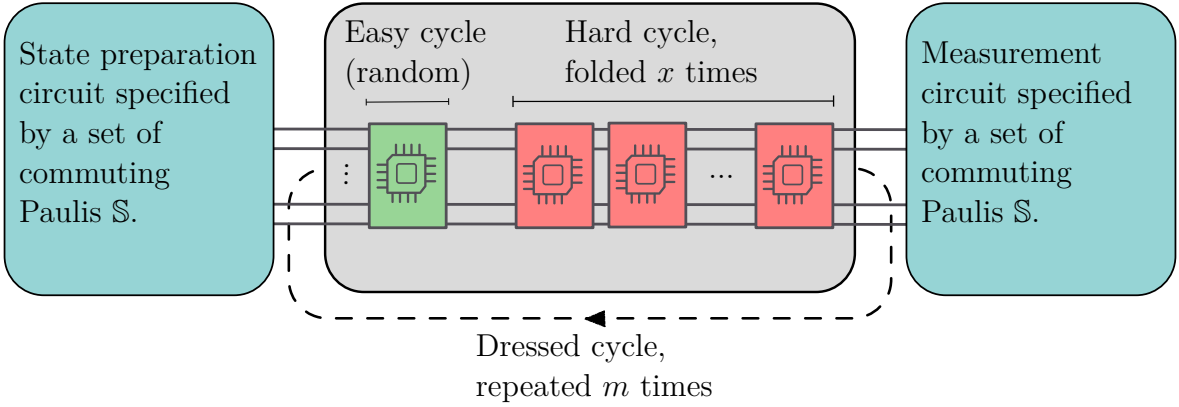
\caption{Visual representation of the CB-structured circuits in our error diagnostic scheme. The core cycles in gray are wrapped in state preparation and measurement (SPAM) circuits specified by a set of commuting Paulis $\set S$. Dressed cycles are repeated $m$ times and they each contain the $x$-folded hard cycle of interest {$(\mc{E}_\hard \circ C_\hard)^x$}, where $C_\hard^x = C_\hard$. The easy cycles are randomized (the random aspect of the circuit is specified by a randomly sampled string $s$) but chosen together with $m$ and the SPAM circuits such that the whole circuit amounts to a Pauli (which is accounted for in post-processing).
}
\label{fig:circuit_representation}
\end{figure}

To distinguish coherent error contributions from decoherent ones, we propose to perform CER sequences on folded hard cycles $(\mc E_\hard \circ C_\hard)^x$ {for $x$ such that $C_\hard^x = C_\hard$.} The circuit representation of the experiments can be visualized in \cref{fig:circuit_representation}. {As such, the pseudo-code for this generalization is almost identical to CER.} 

This idea is not generally equivalent to considering $\mc E_\hard^x \circ C_\hard$, but as we shall see, {they are very closely related in many physically realistic scenarios}. 
Let $c$ be the smallest {positive} integer such that $C_\hard^c=I$, also referred to as the {period} of $C_\hard$. For instance, if the hard cycle consists of parallel CNOTs, we get $c=2$ since CNOT is self-inverting. Without loss of generality, let's express $\mc E_\hard$ as
\begin{align}\label{eq:phase_comm_exp}
    \mc E_\hard =: \exp \left( \Lambda_\hard\right)  =:\exp\left({\Lambda_\hard^{(0)}+\Lambda_\hard^{(1)}+ \cdots \Lambda_\hard^{(c-1)}}\right)~,
\end{align}
where the components $\Lambda_\hard^{(j)}$ phase-commute with $C_\hard$ according to the following equation:
\begin{align}
    \Lambda_\hard^{(j)} \circ C_\hard = e^{\frac{j}{c} 2\pi i} \cdot C_\hard \circ \Lambda_\hard^{(j)}~.
\end{align}
Now, let's consider multiplying the {noisy hard cycle $\mc E_\hard \circ C_\hard $ with itself a number of $c$ times where $c$ is the period of $C_\hard$:
\begin{align}
    (\mc E_\hard \circ C_\hard )^{c} =: e^{\Lambda_{\echo}}~,
\end{align}
where $\Lambda_{\echo}$ is a transformation of the original Lindbladian $\Lambda_\hard$. 
From the Baker-Campbell-Hausdorff formula \cite{BCH}, we get
\begin{align}
    \Lambda_{\echo} &= \sum_{j=1}^c C_\hard^{-j} \circ \Lambda_\hard \circ C_\hard^j +  \frac{1}{2}\sum_{j=2}^{c} \sum_{k = 1}^{j-1} \Bigg[  C_\hard^{-j} \circ \Lambda_\hard \circ C_\hard^{j} , 
C_\hard^{-k} \circ \Lambda_\hard \circ C_\hard^{k}\Bigg] 
+ \cdots \notag \\ 
& = \sum_{k = 0}^{c-1}\sum_{j=1}^c  e^{\frac{k j}{c}2 \pi i} \Lambda_\hard^{(k)} +O(c^2\Lambda_\hard^2) \notag \\
& = c \Lambda_\hard^{(0)} +O(c^2\Lambda_\hard^2) \label{eq:comm_build}
\end{align}
In other words, up to second-order corrections, 
the Lindbladian of the echoed transformation $\Lambda_{\echo}$ 
is the period $c$ times the component of the Lindbladian that commutes
with the hard cycle $C_{\hard}$. 
Therefore, \cref{eq:comm_build} essentially states that by folding the noisy cycle $\mc E_\hard \circ C_\hard$, we effectively evolve the component of $\mc E_\hard$ that commutes with $C_\hard$. This often consists of the only substantial error component. Indeed, $\mc E_\hard \circ C_\hard = C_\hard \circ \mc E_\hard$ is usually a good approximation if the hard cycle of interest $C_\hard$ consists of a round of parallel pulses, since weak non-commuting noise components are expected to be echoed out by the drive, similarly as in \cref{eq:comm_build}. 
}
{
By performing CER sequences on folded hard cycles $(\mc E_\hard \circ C_\hard)^x$ where $C_\hard^x = C_\hard$, we effectively get an experiment to obtain the following fidelities using \cref{eq:comm_build}.
\begin{align}\label{eq:cer_xfid_0}
     f_P^{\rm CER}(x) =& f_P\left(\avg{\mc E_{Q}}_{Q \in \set P_n} \circ (\mc E_\hard \circ C_\hard)^x \circ C_\hard^{-1}\right)~, \notag \\
      =& f_P\left(  \exp\left(x\Lambda_\hard^{(0)}+(\Lambda_\hard-\Lambda_\hard^{(0)}) + \Lambda_Q + O(x^2 \Lambda_\hard^2+ x \Lambda_\hard \Lambda_Q) \right)\right)  \notag \\
     \simeq & f_P\left(  \exp\left(x\Lambda_\hard^{(0)}+(\Lambda_\hard-\Lambda_\hard^{(0)}) + \Lambda_Q \right)\right) ~,
\end{align}
where $\Lambda_Q := \log (\avg{\mc E_{Q}}_{Q \in \set P_n})$ and where the second line is obtained from the first-order Baker-Cambell-Hausdorff formula \cite{BCH}.  
Standard CER fidelities from \cref{eq:cer_fid} are obtained by choosing $x=1$.
In the scenario where $\mc E_\hard \circ C_\hard = C_\hard \circ \mc E_\hard$ (equivalently if $\Lambda_\hard = \Lambda_\hard^{(0)}$)
and where the easy cycle has a very small error component compared to the hard cycle ($\|\Lambda_Q\| \ll \|\Lambda_\hard\|$), we immediately get:
\begin{align}
     f_P^{\rm CER}(x) 
     \simeq & f_P\left(  \exp\left(x\Lambda_\hard^{(0)}\right)\right) ~.
\end{align}
More generally, we get 
\begin{align}\label{eq:cer_xfid}
     f_P^{\rm CER}(x) = f_P\left(  \exp\left(x\Lambda_\hard^{(0)}\right) \right) + \epsilon(x) ~,
\end{align}
where the small term $\epsilon(x)$ is implicitly defined and can approximately take the form of a constant plus a linear term in $x$ if we Taylor expand \cref{eq:cer_xfid_0}. In terms of physics, the linear term in $x$ in the expansion of $\epsilon(x)$ can arise if the coherent part of $(\Lambda_\hard-\Lambda_\hard^{(0)}) + \Lambda_Q$ coherently interferes with the folded Lindbladian $x\Lambda_\hard^{(0)}$.}
{Combining \cref{eq:cer_xfid} with the results of \cref{sec:theory-section,subsec:cer}, we get that by performing CER sequences on folded hard cycles $(\mc E_\hard \circ C_\hard)^x$ where $C_\hard^x = C_\hard$, we should expect decay formulas of the form:

\begin{align}
\label{eq:CER_x_decay}
 \mbb E_s f^{\rm circuit}_P(m , x , s) = A_P \left(1-\delta f^{\rm coh.}_Px^2 - \delta f^{\rm decoh.}_P x+\epsilon(x)\right)^m~,
\end{align}
where the coherent and decoherent Pauli infidelities $\delta f^{\rm coh.}_P$ and $\delta f^{\rm decoh.}_P$
are defined in \cref{eq:infidelities},
but are to be associated with the $\Lambda_\hard^{(0)}$ effective Lindbladian that commutes with $C_\hard$, and where $\epsilon(x)$ is a linear function of $x$. The
impact of $\epsilon(x)$ on the estimates of the decoherent term is expected to be negligible in the regime where the decoherence induced by the hard cycle dominates over the coherent interference between $(\Lambda_\hard-\Lambda_\hard^{(0)}) + \Lambda_Q$ and $x\Lambda_\hard^{(0)}$. 
}

Finally, the coherent contributions $e_P^{\rm coh.}$ to the error probabilities $e_P(\mc E_\hard)$ of the hard cycle $C_\hard$ are obtained by gathering the quadratic components from \cref{eq:CER_x_decay} and by performing the Walsh-Hadamard transform.

\section{Extracting Coherent and Decoherent Qubit Errors Using CER}\label{sec:setup}
\label{sec:Magnon-example}

In this section, we elaborate on a simple example of the CER extension protocol introduced in \cref{subsec:CER-error}. This allows us 
to expand further on the subtleties of the fitting model such as the properties of the underlying covariance matrix. Finally, we provide numerical evidence that the error profile obtained from our suggested experiment matches the underlying error model.

{
We frame our analysis and fitting procedure around a simple proof-of-concept experiment for the sake of concreteness and clarity.
In particular, we focus on the 
errors occurring on an idling qubit during a hard cycle where the nearby qubits undergo a CNOT gate.
We focus on this example for two reasons. First, we judged that simplicity would allow us to expand more explicitly on the fitting procedure and analysis. Second, based on the physics behind IBM processors, we expected (and observed) a noticeable coherent $Z$ error on the idling qubit resting next to the CNOT operation \cite{Magesan_2020}. This highlights the relevance of our method for the characterization of coherent crosstalk effects. All that said, we emphasize once more that our method extends beyond the characterization of a single qubit because it consists of a straightforward generalization of CER, and that CER has been demonstrated on multi-qubit hard cycles \cite{CER_2023,cer_aps}.}

\subsection{Fitting Model}
\label{subsec:fit-model}

\begin{figure}
    \centering
    \includegraphics{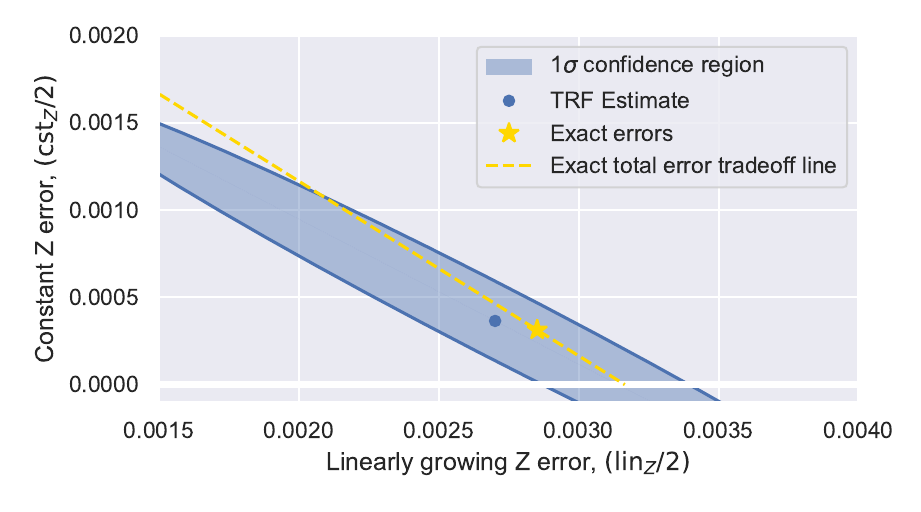}
    \caption{The blue cropped ellipse represents the $1\sigma$ confidence region of 2 fitted parameters, namely $\text{lin}_Z/2$ and $\text{cst}_P/2$, based on the covariance matrix returned by the fitting function. These $2$ parameters appear in the $12$-parameter model given by \cref{eq:twelve parameters_app}. Recall that those parameters are physically constrained to be positive, hence the cropping of the region. The fit was performed on simulated data in order to include exact values (see \cref{sec:fitting}). 
    The solid blue dot is the value returned by the fit, and the star is the exact value of the parameter pair. The dotted line {represents} a tradeoff line where $(\text{lin}_Z+\text{cst}_P)/2$ is set to the exact value, for different values of $\text{lin}_Z$ and $\text{cst}_Z$. The longer principal axis of the ellipse in the figure is  nearly aligned with the dotted tradeoff line. This illustrates that the estimate of $(\text{lin}_Z+\text{cst}_P)/2$ is much more precise than the estimate of $(\text{lin}_Z-\text{cst}_P)/2$.}
    \label{fig:anticorrelation}
\end{figure}

\begin{figure}
    \centering
    \includegraphics{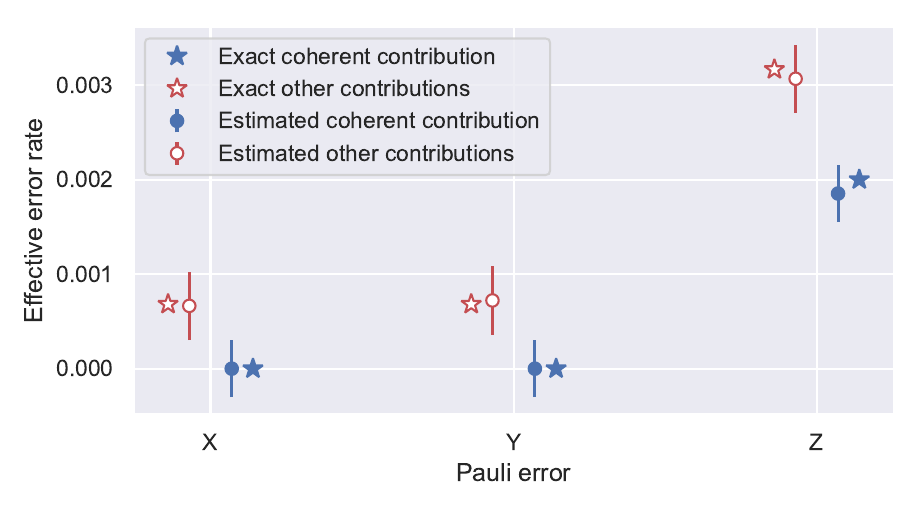}
    \caption{Error profile obtained from a simulation of our experiment. {The coherent contributions refer to the hard cycle coherent errors.} For the simulated error model, we used the relaxation times from \texttt{ibmq\_montreal}, and introduced a coherent $Z$ error with effective rate of $0.002$.}
    \label{fig:simulation_budget}
\end{figure}

As discussed in \cref{subsec:CER-error}, our experiment consists of sampling CER circuits for 
various circuit parameter tuples $(x,m,\set B, s)$:
\begin{itemize}
    \item $x$ is the number of folded  hard cycle per dressed cycle;
    \item $m$ is the number of dressed cycles ;
    \item $\set B$ denotes a {state preparation and measurement} (SPAM) basis as a set of commuting Pauli observables.  We omit $\set B$ in the expression of the Pauli fidelities since it is {implicitly} chosen through the $P$ Pauli index.
    \item $s$ denotes a randomly sampled string that encodes the random part of the circuit construction for the fixed parameters $(x,m,\set B)$.
\end{itemize}
From those parameterized sampled circuits, we generated fidelity-like quantities $f^{\rm circuit}_P(x, m , s)$.  Consider the simplest instance of such an experiment on a single qubit.  According to \cref{eq:CER_x_decay}, the sample average of $f^{\rm circuit}_P(x, m , s)$ obeys the following $12$-parameter decay model:
\begin{equation}
\label{eq:twelve parameters_app}
\begin{split}
  \mbb E_s f^{\rm circuit}_X(x, m , s) =& A_{X}\Big(1-(\text{quad}_{Y}+\text{quad}_{Z})x^{2}-(\text{lin}_{Y}+\text{lin}_{Z})x-(\text{cst}_{Y}+\text{cst}_{Z})\Big)^m  \\
  \mbb E_s f^{\rm circuit}_Y(x, m , s) =& A_{Y}\Big(1-(\text{quad}_{X}+\text{quad}_{Z})x^{2}-(\text{lin}_{X}+\text{lin}_{Z})x-(\text{cst}_{X}+\text{cst}_{Z})\Big)^m   \\
  \mbb E_s f^{\rm circuit}_Z(x, m , s) =& A_{Z}\Big(1-(\text{quad}_{X}+\text{quad}_{Y})x^{2}-(\text{lin}_{X}+\text{lin}_{Y})x-(\text{cst}_{X}+\text{cst}_{Y})\Big)^m 
\end{split}
\end{equation}
{There are many possible methods to retrieve estimates for the $12$ parameters
($A_X$,$A_Y$, $A_Z$, $\text{quad}_{X}$, $\text{quad}_{Y}$,$\text{quad}_{Z}$, $\text{lin}_{X}$,$\text{lin}_{Y}$,$\text{lin}_{Z}$,$\text{cst}_{X}$, $\text{cst}_{Y}$,$\text{cst}_{Z}$ ). A simple one is to start with an initial guess and perform a minimization algorithm with the cost function defined
by 
\begin{align}\label{eq:chi_model}
\chi^2(A_X,A_Y, A_Z, \text{quad}_{X}, \text{quad}_{Y},\text{quad}_{Z}, \text{lin}_{X},\text{lin}_{Y},\text{lin}_{Z},\text{cst}_{X}, \text{cst}_{Y},\text{cst}_{Z} ) : =\notag \\
 \sum_{P, m,x} \left( \mbb E_s f^{\rm circuit}_P(x, m , s) - A_P \left(1-\sum_{\substack{Q\\ QP=-PQ}} (\text{quad}_Q x^2 + \text{lin}_Q x +\text{cst}_Q)\right)^m \right)^2~,
\end{align}
where the sum $\sum_{P, m,x}$ is performed over the desired Paulis, sequence lengths and number of folded hard cycles. For instance, in this work, 
we used the Trust-Region-Reflective (TRF) least mean squares algorithm to fit the parameters to the fidelity data \cite{2020SciPy}.}
In the general case, for $N$ different Pauli fidelities to estimate, the number of parameters in the fit becomes $4N$. Since CER focuses on error probabilities marginalized over a small number of qubits, the number of parameters is guaranteed to be manageable unless we observe long-range error correlations.

As shown in the previous section, the coefficients $\text{quad}_P$ appearing in the quadratic part of the exponential decay are in direct correspondence with the effective coherent contribution to the infidelity of the hard cycle on the qubit:
\begin{align}\label{eq:coh}
    \text{quad}_P  = 2e_P^{\rm coh.}~.
\end{align}
The linear coefficients are most realistically in correspondence with the decoherent contribution to the infidelity of the hard cycle on the qubit, although some linear effects could in principle be induced by a coherent effect between the randomized easy cycle and the folded hard cycle. Easy cycles are usually much less error-prone than hard cycles, meaning that the linear coefficient is often entirely dominated by the decoherent contribution. Finally, the constant terms $\text{cst}_P$ are due to the errors occurring during the easy cycles, as well as
to the components of the hard cycle error that does not commute with the hard cycle $C_\hard$ (see \cref{eq:cer_xfid}).

\subsection{Resource requirements}

{
CER as well as its newly introduced variant are designed for scaling well with the number of qubits $n$. The number of circuits for such 
experiments scales as the number of marginal error probabilities to extract, and often -- due to constraints in the correlations of errors -- 
only a polynomial number of marginals is enough to accurately describe the error profile. Moreover, other hyper-parameters, 
such as the number of shots and the number of random circuit samples can be kept constant and still yield multiplicative precision
estimates on the error probabilities. Regarding the required number of shots, it is shown in \cite{Harper2019statRB} that
the number of shots can be kept constant no matter the error rates as long as the sequence lengths are appropriately chosen
(the optimal choice of sequence lengths is 
discussed in \cite{Harper2019statRB}). 
It is worth mentioning that the number of shots bounds the relative precision of the error probability estimates, 
so it is worth picking a few thousand shots to ensure 2 significant digits on the estimates. Regarding the number 
of random circuit samples, it is known that due to concentration inequalities, randomly compiled channels converge
exponentially fast to their true average with the number of samples. For instance, in \cite{Goss2023} it is shown 
that 30 circuit samples are enough to heavily suppress coherent errors (see the figure $2$ in the paper).
}

\subsection{Fitting Model Simulation Results and Analysis}
\label{sec:fitting}

To anticipate our hardware-based experiments, we perform numerical simulations of our experiment in the simple case of a single qubit subject to a noise source coming from a nearby entangling gate.  To measure the systematic increase in the magnitude of the error from the noise, the number of CNOT gates is systematically increased.  Measurements are taken for hard gate cycles of  1, 3, 5, 7 and 9 CNOT gates and the simulations are run for 20,000 shots per random circuit. {We used the values 4, 8, 12, 16 and 32 for the the sequence lengths}. For the simulated error model, we use the relaxation times from \texttt{ibmq\_montreal}, and introduce a coherent $Z$ error with an effective rate of $0.002$ to simulate the presence of coherent crosstalk.

In our numerical analysis, we notice that $\text{lin}_P$ and $\text{cst}_P$ are strongly anti-correlated, inducing a large uncertainty for their difference $\text{lin}_P-\text{cst}_P$. This is well illustrated in \cref{fig:anticorrelation}, where we can further see that the sum $\text{lin}_P+\text{cst}_P$ is estimated with high precision and accuracy. For this reason, we contrast the coherent contribution to the error rates {from the hard cycle} with the sum of the other contributions (see \cref{fig:simulation_budget}). In the regime where the twirling operations are nearly perfect, the contributions $\text{lin}_P+\text{cst}_P$ can be interpreted as the sole result of decoherence during the hard cycle. More generally, one could make tighter upper-bounds on the constants $\text{cst}_P$, based on physical assumptions or on additional benchmarking data. For example, in \cref{fig:anticorrelation}, one could reduce the uncertainty on the difference $\text{lin}_P-\text{cst}_P$ by upper-bounding the constant $Z$ error $\text{cst}_Z/2 \leq 0.0005$. This would crop the confidence region and improve the precision and level of detail of the error profile, but would inherently rely on an assumption.

In all of the computations, the error parameters appearing in the model described by \cref{eq:twelve parameters_app} are bounded between $0$ and $1$ (they are usually close to $0$).  As such, we use the Trust-Region-Reflective (TRF) least mean squares algorithm to fit the parameters to the data \cite{2020SciPy}.  We compare the estimated error profile to an exact reference in \cref{fig:simulation_budget} and find a satisfactory degree of agreement.

\section{Hardware results and analysis}
\label{data analysis}

Starting with the proposed fitting model discussed in \cref{subsec:fit-model}, we note that each of the quadratic equations $f_{X}$, $f_{Y}$ and $f_{Z}$ in \cref{eq:twelve parameters_app} are interconnected because each direction (X, Y, Z) has anti-commuting terms for X, Y, and Z that are coupled across each dimension of the model.  This requires each of the coefficients in the polynomial to have both a coherent ($ \rm{quad}_{P}$) and a decoherent ($ \rm{lin}_{P}$) fitting parameter. Having all possible terms that anti-commute with each of the x, y, and z axes in \cref{eq:twelve parameters_app} included in the proposed model require twelve parameters ( 3 $``A_{P}"$ parameters, 3 $``\rm{quad}_{P}"$ parameters, 3 $``\rm{lin}_{P}"$ parameters and finally 3 $``\rm{cst}_{P}"$ parameters).  One advantage of such an approach is that the global fit to these twelve parameters assures overall positive error rates.  The results from this global fitting produce 75 values (five different CNOT repetitions at the five different sequence lengths and three $f_{X}$, $f_{Y}$ and $f_{Z}$).   

We construct graphs of the component fidelities $f_{X}$, $f_{Y}$ and $f_{Z}$ versus sequence lengths for hard cycle CNOT repetitions 1, 3, 5, 7, and 9 for the fitting model for each of the three different hardware platforms.  These 75 values are plotted on graphs of $f_{X}$, $f_{Y}$ and $f_{Z}$ versus sequence lengths for each hard cycle CNOT repetition (1, 3, 5, 7, and 9) with star symbols ($\star$).  ~\cref{fig:fit-Manila} shows an example of these graphs for the \texttt{ibmq\_manila} hardware platform. 

To compare these computations with experimental data we selected the cloud-accessible IBM quantum computing hardware platforms as a proof-of-concept of our experimental design derived in \cref{subsec:CER-error}.  The cycle and marginal distribution selected are modelled from the circuit design implemented to measure the spin-spin correlation function of a Heisenberg spin chain \cite{francis2020quantum}.  \cref{fig:Magnon-circuit} is a block diagram of the spin-spin correlation function circuit showing CNOT gates in the circuit and an ancilla qubit used to measure the computational results. \cref{fig:time_evo} shows a more detailed sub-circuit used in the time evolution of the Heisenberg spin-chain. As shown through the diagrams, the types of circuits that we consider feature a spectator ancillary qubit that is left idling during the trotterized evolution of the spin chain. As such, we focused our interest on the coherent and decoherent error profile marginalized on this ancilla. In this simple instance, the hard cycle of interest consist of a CNOT acting on a pair of qubits  neighboring to the idle ancilla.  These types of circuits motivated this work.

We implement the circuit on three different IBM quantum computing hardware platforms (\texttt{ibmq\_guadalupe, ibmq\_montreal, and ibmq\_manila}). \cref{appendix:IBM-hardware} discusses the specific circuit implementations on each of the hardware platforms.   
\begin{figure}[h]
    \centerline{
\Qcircuit @C=1em @R=.7em {\lstick{\ket{0}} & \gate{H} & \qw & \ctrl{1} \qwx[1] & \qw &\qw & \qw & \ctrl{1} \qwx[1] & \qw & \meter\\
& \multigate{3}{\hat{U}_S} & \qw & \multigate{3}{\hat{U}^A} &\qw & \multigate{3}{e^{-iHt}} & \qw & \multigate{3}{\hat{U}^B} & \qw \\
& \ghost{\hat{U}_S}& \qw &  \ghost{\hat{U}^A}& \qw & \ghost{e^{-iHt}}& \qw & \ghost{\hat{U}^B}  & \qw \\ 
& \ghost{\hat{U}_S} & \qw & \ghost{\hat{U}^A}& \qw & \ghost{e^{-iHt}}& \qw  & \ghost{\hat{U}^B}  & \qw\\ 
& \ghost{\hat{U}_S} & \qw &  \ghost{\hat{U}^A}& \qw & \ghost{e^{-iHt}}& \qw & \ghost{\hat{U}^B}  & \qw
\inputgroupv{2}{5}{1.3em}{2.4em}{\ket{0}^{\otimes n}\;\;} \\
}}
\caption{Block diagram for calculating spin-spin correlation in 4-site Heisenberg spin-chain \cite{francis2020quantum}.}
				\label{fig:Magnon-circuit}
			\end{figure}
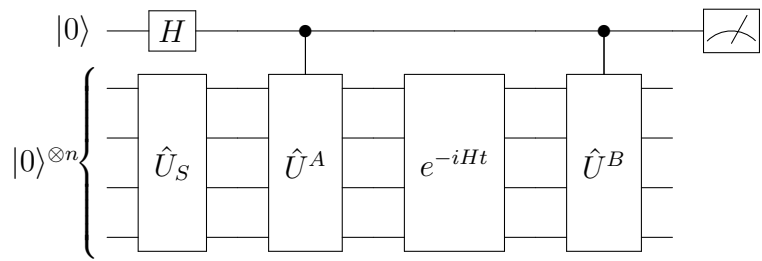
\begin{figure}[h]
    \centerline{
\Qcircuit @C=1em @R=.7em {& \qw & \ctrl{1} \qwx[1] & \gate{R_x\left(-2Jt-\pi/2\right)} & \gate{H} & \ctrl{1} \qwx[1] & \gate{H} & \ctrl{1} \qwx[1] & \gate{R_x(\pi/2)}\\
& \qw & \gate{X}  & \gate{R_z(-2Jt)} & \qw & \gate{X} & \gate{R_z(2Jt)} & \gate{X} & \gate{R_x(-\pi/2)}}}
\caption{Sub-circuit used in time-evolution (see $e^{-iHt}$ block of ~\cref{fig:Magnon-circuit}) of the 4-site Heisenberg spin-chain. {Four instances of this sub-circuit implement $e^{-iHt}$ block of ~\cref{fig:Magnon-circuit}}}
	\label{fig:time_evo}
\end{figure}
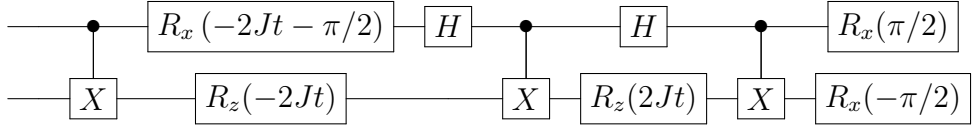
These calculations are randomly compiled 30 times with different Paulis for each sequence length on each of the three hardware platforms using 10,000 shots for \texttt{ibmq\_manila} and 20,000 shots for \texttt{ibmq\_guadalupe} and \texttt{ibmq\_montreal}.  We color-code each of the 30 results from these experimental measurements for each sequence length and plot them on the graph of $f_{X}$, $f_{Y}$ and $f_{Z}$ versus sequence lengths for each hard cycle CNOT repetition (1, 3, 5, 7, and 9) and each of the three different hardware platforms.  We also plot the sample mean from the 30 different data points for each hard cycle CNOT repetition at each sequence length on the graph with the diamond symbol($\diamond$) and a standard deviation error bar.  \cref{fig:fit-Manila} shows these results for the \texttt{ibmq\_manila} hardware platform. We construct similar computations and graphs for both \texttt{ibm\_montreal} and \texttt{ibm\_guadalupe}.  

Upon examination of \cref{fig:fit-Manila} in more detail, it can be seen that some of the 75 different pairs of stars and diamonds do not overlay each other on the graph.  For example, the 3 CNOT folding star-diamond pairs for all of the sequence lengths 4, 8, 12, 16, and 32 in the component fidelity $f_{Y}$ are slightly displaced from each other. Similar examples of these star-diamond pair displacements can also be seen throughout the other CNOT foldings in the figure. {However, these displacements are rather small. We calculated the coefficient of determination, also referred to as $R^2$, \cite{statistic_ref},  to three significant digits for all three devices (for \texttt{ibmq\_manila}, $R^2 = 0.894$; for \texttt{ibmq\_montreal}, $R^2 = 0.989$; for \texttt{ibmq\_guadalupe}, $R^2 = 0.990$). The fact that these values are well above $80\%$ indicates that the model substantially accounts for all the data.}

We then implement a minimization procedure on the star-diamond data.  The goal is to globally minimize the distance between the diamond values and the star values for all 75 star-diamond combinations.  We perform this global minimization by computing the square of the difference between each pair of stars and diamonds and dividing by that specific star value and then summing over all 75 star/diamond pairs   This calculation is essentially {a minimization of the $\chi^{2}$ statistic to improve the goodness of the fit.

From \cref{sec:fitting} a more detailed analysis for the 12-parameter fit shows that the sums and differences of the $\text{lin}_P/2$ and $\text{cst}_P/2$ are strongly anti-correlated. As stated in \cref{eq:coh}, the $\text{quad}_P/2$ values contain the coherent contribution to the error rate {from the hard cycle}, while the $(\text{lin}_P+\text{cst}_P)/2$ contains the other contributions. We plot these terms for each of the three IBM hardware platforms as shown in \cref{fig:Effective Error Rates All}. This figure is the central result of this research project and shows the effective error rate versus Pauli Error (X, Y and Z) for the designated single qubit on each of the hardware platforms. These results as well as the data from the simulator are shown in \cref{Tab:2-eff error rate}. The table clearly shows that the  Pauli Z $\text{lin}_P/2$ and $\text{cst}_P/2$ terms have the strongest error measured on the single qubit on each of the hardware platforms.

\begin{figure}[!htpb]
    \centering
    \includegraphics{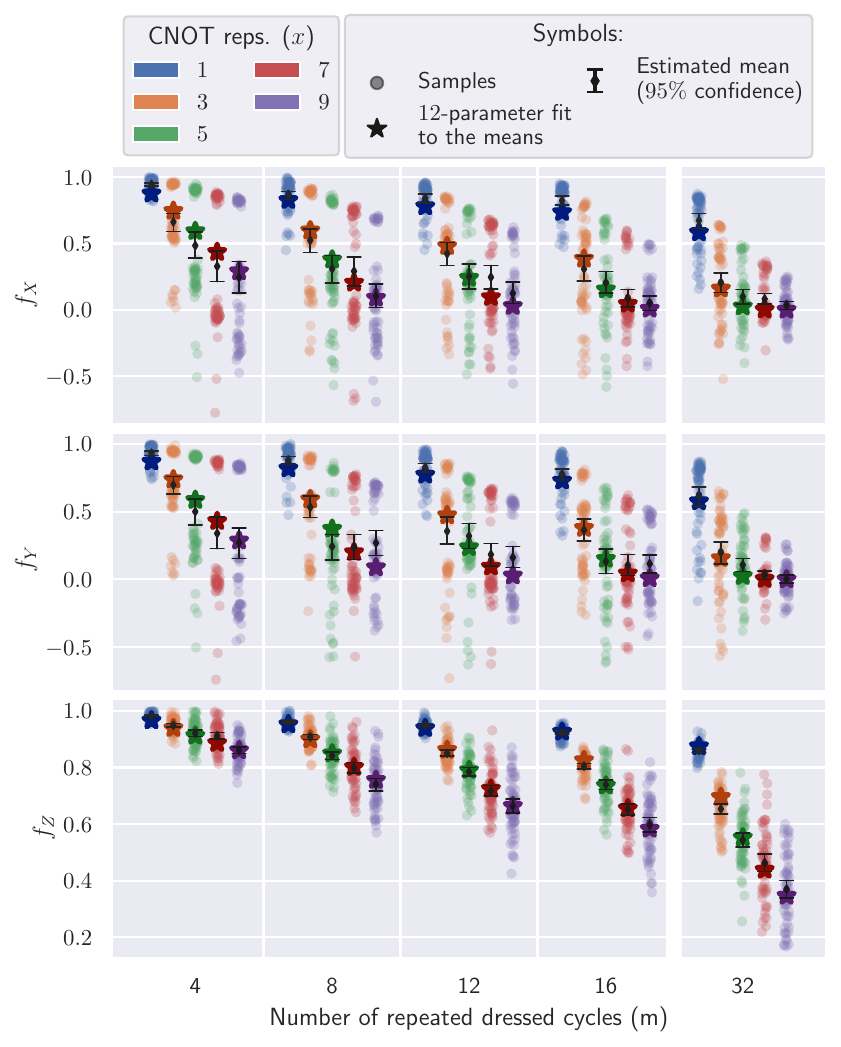}
\caption{ Fidelities $f_{X}$, $f_{Y}$ and $f_{Z}$ versus sequence lengths for CNOT repetition 1,3,5,7, and 9 for \texttt{ibmq\_manila}. Each CNOT repetition is color coded and identified in the legend in the upper left portion of the figure. Each CNOT repetition was randomly compiled 30 times with different Paulis at each sequence length.  Each individual data point plotted is color code matched to the corresponding CNOT repetition. The average value from the 30 different data points for each CNOT repetition at each sequence length is then plotted on the graph with the diamond symbol($\diamond$) and standard deviation error bar. The global fitting from the model produces 75 values for the estimated mean and is plotted with star symbols ($\star$)}
\label{fig:fit-Manila}
\end{figure}

\begin{figure}[!htpb]
    \centering
    \includegraphics{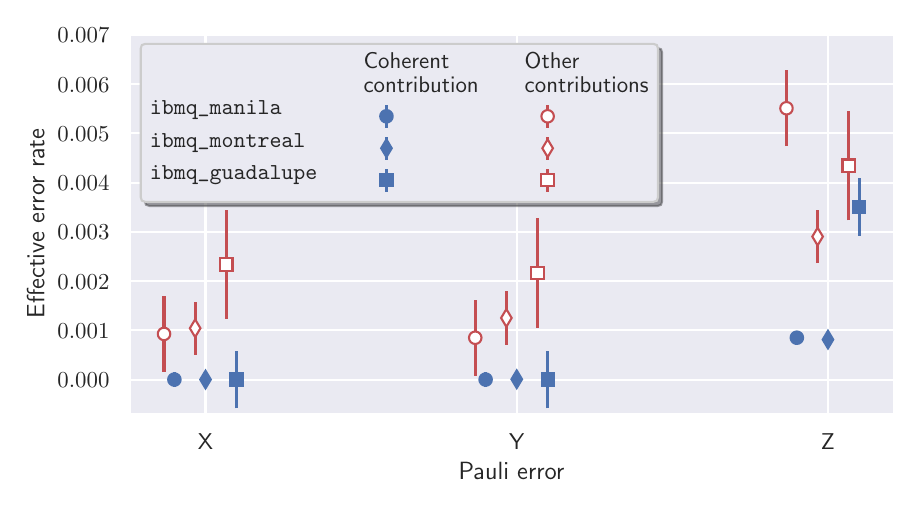}
\caption{Effective error profile for an idling ancilla standing next to a CNOT on IBM hardware platforms \texttt{ibm\_guadalupe}, \texttt{ibm\_montreal} and \texttt{ibm\_manila}. The horizontal axis contains the three possible Pauli errors on the ancilla, and the vertical axis quantifies the respective effective error probability (when rendering the error channel Pauli stochastic through a projection). The solid blue markers indicate the coherent contribution to the error rate {from the hard cycle}, while the red open markers quantify the other contributions to the error rate. The results are obtained via the extended CER protocol described in \cref{sec:CER protocol}. The numerical values of the error rates with their respective uncertainty are contained in \cref{Tab:2-eff error rate}.}
\label{fig:Effective Error Rates All}
\end{figure}

It is well-known that the static $ZZ$ coupling in a transmon qubit is always present and leads to both coherent and incoherent errors \cite{tripathi2108suppression}. Recent works have theoretically analyzed the effect of crosstalk on simultaneous gate operation in a tunable ZZ coupling-based qubit architecture \cite{zhao2022quantum,long2021universal}. Our results confirmed that a CNOT gate indeed affects the idling qubit, and the nature of the coherent error is predominantly Pauli $Z$ error. We observed this noise bias toward Pauli $Z$ error due to the effects of static $ZZ$ coupling in a transmon qubit. 

This increase in the single qubit error as the number of CNOT hard cycles is increased can also be seen graphically through heat map plots for the single qubit on \texttt{ibm\_guadalupe}, \texttt{ibm\_montreal} and \texttt{ibm\_manila} as shown in \cref{fig:guad_heatmap}, \cref{fig:montreal_heatmap} and \cref{fig:manila_heatmap} in \cref{appendix:CER-Heat}.  Although there is some increase in the $X$ and $Y$ Pauli error, these heatmaps also graphically illustrate that it is the Z Pauli error that shows the greatest increase as the number of CNOT hard cycles increases.  We observe from our heatmaps that Pauli $Z$ error increases rapidly compared to $X$ and $Y$ errors when the  number of CNOTs increases from 1 to 9. This rapid scaling of $Z$ error is the signature of coherent errors. Thus, our characterization scheme shows not only that  Pauli $Z$ error is the dominant error on the  IBM quantum processors, but also that a substantial fraction of it is the result of coherent processes.

\section{Summary}
\label{sec:Summary}

We successfully demonstrate the design and implementation of an efficient and scalable diagnostic method that quantitatively differentiates between coherent and decoherent errors in cycles. The characterization scheme that we suggest in the present work differs from existing error diagnostic methods such as GST \cite{Merkel2013,Blume-Kohout2013,Blume_Kohout_2017,Greenbaum2015, Nielsen_2021,Brieger2021} in that it is targeted toward 
the characterization of the effective dressed cycles present in randomly compiled circuits. 
This is an important tool because many error suppression techniques implemented today only provide improved circuit performance by focusing on mitigating either decoherent or coherent errors. Our method can therefore be used to measure the impact of error suppression suites on the type of errors that they specifically target.

We leverage a CB-structured error characterization protocol known as CER \cite{erhard2019characterizing,CER_2023,cer_aps}. The original method was designed to measure the error profile on effective dressed cycles, which are tailored to have purely decoherent errors via a compiling method known as RC \cite{wallman2014randomized}. We expand on the basic CER method by introducing an additional hyper-parameter (labeled $x$ in this work) which corresponds to the number of hard cycle repetitions before being subject to Pauli twirling (see \cref{fig:circuit_representation}). This additional hyper-parameter allows for quantitative estimates of the coherent error contributions to be computed through a generalization of the fidelity decay formula (\cref{eq:CER_x_decay}).  

Our data analysis relies on the different propagation formulas of coherent and decoherent errors in folded error channels (see \cref{sec:theory-section}). As a proof of concept, we test our method both physically and numerically by reconstructing the effective error profile on a single-qubit ancilla left idling during a cycle. The numerical simulation confirms a strong agreement between the error profile estimated by our protocol and the exact underlying error model. The data obtained from IBM hardware platforms (\texttt{ibmq\_guadalupe}, \texttt{ibmq\_manila}, and \texttt{ibmq\_montreal}) revealed a substantial level of coherent errors occurring on the idling ancilla induced by an entangling operation (see \cref{fig:Effective Error Rates All}).

In terms of next steps, we note that the fitting method derived in the current work only relies on the sample averages of fidelities, and does not take into consideration the shape of the distribution of the fidelities for fixed $(m,x)$ hyper-parameters. As shown in \cref{fig:fit-Manila}, coherent errors affect more properties of the fidelity distributions than just the mean and we leave as an open problem the refining of the fitting function based on those considerations, as well as statistical optimization of the choice of hyper-parameters. Finally, we leave the demonstration of our proposed CB-structured method as a means to design and benchmark error suppression tools for future work.


    

\begin{table}[h!]
  \begin{center}
  \caption{Various fitted parameters corresponding to the $12$-parameter model presented in \cref{eq:twelve parameters_app}, for three different hardware platforms. We include the difference and the sum of $\text{lin}_P/2$ and $\text{cst}_P/2$ since these parameters are strongly anti-correlated (see \cref{sec:fitting}). From \cref{eq:coh}, we get that the $\text{quad}_P/2$ column contains the coherent contribution to the error rate {from the hard cycle}, while the $(\text{lin}_P+\text{cst}_P)/2$ contains the other contributions. These error rates are shown in \cref{fig:Effective Error Rates All}.}
    \label{Tab:2-eff error rate}
    \begin{tabular}{|l|l|l|l|l|l|}
     \hline 
       Pauli error &  $\text{quad}_P/2$ & $\text{lin}_P/2$ & $\text{cst}_P/2$ & $(\text{lin}_P+\text{cst}_P)/2$ & $(\text{lin}_P-\text{cst}_P)/2$\\ 
      \hline\hline
 Montreal & & & & & \\
      $X$    & 0.00000(9)  &  0.008(2)   &  0.0004(9) &  0.0010(5) & 0.000(1) \\ 
      $Y$    & 0.00001(9)  &  0.007(2)   &  0.0000(9) &  0.0013(5) & 0.001(1) \\ 
      $Z$    & 0.00081(9)  &  0.0038(7)  &  0.0000(9) &  0.0029(5) & 0.003(1) \\ 
      \hline\hline
Guadalupe & & & & &  \\
      $X$    & 0.0000(6)  &  0.001(3)  &  0.001(3)   & 0.002(1)  & 0.000(6)  \\ 
      $Y$    & 0.0000(6)  &  0.002(3)  &  0.000(3)   & 0.002(1)  & 0.002(6)  \\ 
      $Z$    & 0.0035(6)  &  0.004(3)  &  0.000(3)   & 0.004(1)  & 0.004(6)  \\ 
      \hline\hline
Manila & & & & &  \\

      $X$    & 0.0000(1) &  0.001(1) &  0.000(1) &  0.0009(8) & 0.001(2)\\ 
      $Y$    & 0.0000(1) &  0.001(1) &  0.000(1) &  0.0008(8) & 0.001(2) \\ 
      $Z$    & 0.0008(1) &  0.006(1) &  0.000(1) &  0.0055(8) & 0.006(2)\\ 
            \hline\hline
Simulator & & & & &  \\

      $X$    & 0.0000(3) &  0.000(2) &  0.000(1) &  0.0007(4) & 0.000(3)\\ 
      $Y$    & 0.0000(3) &  0.001(2) &  0.000(1) &  0.0007(4) & 0.001(3) \\ 
      $Z$    & 0.0019(1) &  0.003(2) &  0.000(1) &  0.0031(4) & 0.002(3)\\  
      \hline
    \end{tabular}
  \end{center}
\end{table}

\section{Acknowledgments}

P.D. was supported in part by the U.S. Department of Energy (DoE) under award DE-AC05-00OR22725. S.K.R. acknowledges financial support from the J. William Fulbright Foreign Scholarship Board and the Fulbright Commission in India (USIEF) through a Fulbright Nehru Doctoral Research Fellowship 2021-2022. We acknowledge the use of IBM Quantum services for this work. The views expressed are those of the authors, and do not reflect the official policy or position of IBM or the IBM Quantum team.  We thank North Carolina State University (NCSU) for access to the IBM Quantum Network quantum computing hardware platforms through the NCSU IBM Quantum Hub.  The authors also thank the reviewers for many helpful comments that have improved the final manuscript.  The project team acknowledges the use of True-Q software from Keysight Technologies.

 \section{Author Contributions}
 All authors contributed equally to this project.

 \section{Competing interests}
A. C-D. has a financial interest in Keysight Technology Inc. and the use of True-Q software. P.D. and S.K.R. declare no competing interests.
 
 \section{Data and Code Availability}
The code and data are available from the corresponding author upon reasonable request.

\bibliographystyle{unsrtnat}
\bibliography{ref}
\onecolumn

\appendix

\section{Supplementary Material}

In this section, we prove \cref{lem:fid_quad}, which yields the propagation formula given in \cref{eq:fidel-repeated-channel}.

\subsection{Preliminary definitions}
Before the proof, consider the following definition:
\begin{defn}[Area of Effect]
Consider a connectivity graph $G = (V,E)$, where $V$ is the set of qubit vertices and $E$ is a set of connectivity edges. 
We say that, given the graph $G$, an operator $M$ has an area of effect
\begin{align}
    \aoe{M|G}:= \min_{W \subseteq V} |W|
\end{align}
such that 
\begin{enumerate}
    \item The vertex-induced subgraph obtained by limiting the vertices of G to the set $W$ is connected;
    \item $M$ can be expressed as $M_{H_W} \otimes I_{H_{W^c}}$, where $W^c$ is the complement of $W$, and $H_W$ ($H_{W^c}$) denotes the Hilbert space over the qubits in the set $W$ (${W^c}$).
\end{enumerate}
\end{defn}
An operator $M$ is said to be geometrically (or topologically) $k$-local if $A(M|G) = k$. See \cref{fig:lemma}a for an example of an operator with area of effect of $7$.  As we show, if we limit the area of effect of noisy interactions, early-truncated Taylor expansion of $e^{\Lambda}$
will be appropriate for approximating the effect of $e^{\Lambda }$ on reasonably small subsystems.

{
Let's further consider the extended support of a Pauli basis element:
\begin{defn}[$k$-extended support of a Pauli basis element]
Consider a Pauli basis element $P = P_1P_2 \cdots P_n \in \set P_n$ and a connectivity graph $G = (V,E)$, where $V$ is the set of qubit vertices and $E$ is a set of connectivity edges. The ($0$-extended) support of $P$ is defined as the set of qubits for which $P$ is not the identity:
\begin{align}
    \set S_0(P) := \{ i \in V | P_i \neq \mbb I \}~.
\end{align}
The $k$-extended support of $P$ is simply an extension of the support of $P$ by $k \in \set N$ vertices:
\begin{align}
    \set S_k(P) := \{ i \in V | \exists j \in \set S_0(P) ~{\rm s.t.}~ d(i,j) \leq k \}~,
\end{align}
where $d(\cdot, \cdot)$ is the distance on the graph $G$. 
\end{defn}
See \cref{fig:lemma}b for an example of different extended supports with different values of $k$. Let's define the decoherent error rate over a set of qubits.
}
{
Let's define the decoherent error rate over a set of qubits.
\begin{defn}[decoherent error rate over a set of qubits]
Consider a set of qubits $V$. We define the decoherent 
error rate over those qubits as 
\begin{align}
    e_V^{\rm decoh} &:=  \sum_{\substack{P \\ \set S(P)= V}} e_P^{\rm decoh} \\
    &:= \sum_{\substack{P \\ \set S(P)= V}}\sum_{j} |\ell_{j,P}|^2~.
\end{align}
\end{defn}
}

{
\begin{defn}
Let's define the coherent and decoherent infidelities of $P \in \set P_n$ as:
\begin{subequations}
\begin{align}
    \delta f_P^{\rm decoh.}:= 2\sum_{\substack{j, Q\\ QP=-PQ}} |\ell_{j,Q}|^2\\
    \delta f_P^{\rm coh.}:= 2\sum_{\substack{Q\\ QP=-PQ}}|h_{Q}|^2
\end{align}
\end{subequations}
\end{defn}
}

\begin{figure}[h!]
\centering
\def\svgwidth{\columnwidth}
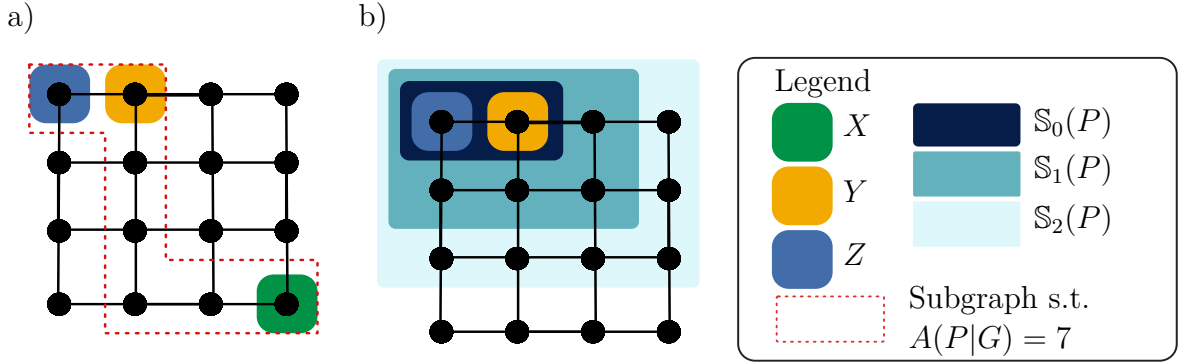
\caption{Visualization of different Paulis $P \in \set P_{16}$ in a planar 16-qubit architecture. a) Visualization of subgraph associated with the area of effect $A(P|G)$ on a planar architecture. The Pauli in figure a) has an area of effect of $7$. b) Visualization of $k$-extended supports of a Pauli for $k \in \{0,1,2\}$.
}
\label{fig:lemma}
\end{figure}

\subsection{Derivation of the Pauli Fidelities of a Repeated Channel}

\label{sec:appendix A} 

\begin{boxlem}{Pauli fidelities of repeated channel}{fid_quad}
   Let Lindbladian and Hamiltonian operators be at most geometrically 
   $k$-local for some integer $k$.
   Consider an error channel $\mc E= e^{\Lambda}$ 
   stemming from geometrically $k$-local interactions.
   The fidelity of $P$ associated with the channel 
   $\mc{E}^x =e^{x\Lambda}$ is given by
\begin{align} 
       f_P(x) 
        = &1-  \delta f_P^{\rm coh.} x^2 - \delta f_P^{\rm decoh.} x \notag \\
        &+O\left({\rm poly}(e^k)\right) \Bigg[ (\delta f_P^{\rm decoh.}) \max_{S \in \set S_k(P)} \left( e_S^{\rm decoh.}\right) \notag \\
        & + \left(2\max_{\substack{S, ~\chi_{S,P}=-1 }}  \left|h_{S}  \right| + \max_{\substack{S \in \set S_k(P)}}  e_S^{\rm decoh.} \right) \times  e_{\set S_k(P)}^{\rm decoh}\Bigg]x^2 \notag \\
       &+\tr \left (P \frac{\Lambda^3}{3!}[P] \right) x^3+ \tr \left (P \frac{\Lambda^4}{4!}[P] \right) x^4 + \cdots.
\end{align} 
\end{boxlem}

\begin{proof}
 A way to interpret the total evolution $e^\Lambda$ is to consider $\Lambda$ as a transition matrix.
 The total evolution depicted by $e^\Lambda$ is then the sum over all paths, and each path is weighted
 by $1/J!$ where $J$ is the number of jumps. 
 The {essence} of the proof will be to quantify and categorize the transition amplitudes, 
 and to sum up the different paths. For conciseness, we omit the graph dependence in the area of effect function, i.e. $\aoe{M|G}$ is replaced by $\aoe{M}$.
\\


 Let's define the transition amplitude from the Pauli $P \in \set P_n$ to the Pauli $Q \in \set P_n$ as
 \begin{align}
    \trans{P}{Q} := \frac{\tr \left( Q \Lambda[P] \right)}{2^n}~.
 \end{align}
 From the definition of the Lindblad matrix \cref{eq:lambda}, we more specifically get 
  \begin{align}\label{eq:trans_cap}
    2^n \cdot \trans{P}{Q} = -i\tr \left( Q [H,P] \right) 
                        + \sum_j \left\{ \tr \left( Q L_jPL_j^\dagger \right)
                                                    -\frac{1}{2}\tr \left( Q PL_j^\dagger L_j  \right)
                                                    -\frac{1}{2}\tr \left( QL_j^\dagger L_j  P \right)\right\}~.
 \end{align}
It follows by the hermicity-preserving nature of the evolution that $\trans{P}{Q}$ remains real. 
Let's decompose the operators $L_{j}$ \cref{eq:a}, $L_j^\dagger$ \cref{eq:b} and $H$ \cref{eq:c} by inserting summations over the Pauli operators $S$ as follows:
\begin{subequations}
    \begin{align}
        L_j &= \sum_S \ell_{j,S} S~, \label{eq:a} \\
        L_j^\dagger &= \sum_S \ell_{j,PQS}^* PQS~, \label{eq:b}\\
        H &= \sum_S h_{PS} PS~, \label{eq:c} 
    \end{align}
\end{subequations}
where
\begin{subequations}
    \begin{align}
        \ell_{j,S}   &:= \tr  (S L_j)/ 2^n~, \\
        \ell_{j,PQS} &:= \tr  (PQS L_j)/ 2^n~, \\
        h_{PS} &:= \tr ((PS)^\dagger H) /2^n~,
    \end{align}
\end{subequations}
and let's substitute those decompositions into
\cref{eq:trans_cap}:
\begin{align}
    \trans{P}{Q} = & -i h_{PQ}\tr \left( Q PQ P - Q P PQ \right)/2^n 
            + \sum_{j,S}  \ell_{j,S} \ell_{j,PQS}^*\tr \left( Q SPPQS \right)/2^n\notag \\ 
            &
        -\frac{1}{2} \sum_{j,S} \ell_{j,S} \ell_{j,PQS}^* \tr \left( (PQ+QP) PQS S  \right)/2^n~.
\end{align}
The above expression can be further simplified. Let's define
the commutation function, 
\begin{align}
    \chi_{P,Q} = \begin{cases}
       \text{$~~1$ if $P$ and $Q$ commute}~, \\
        \text{$-1$ if $P$ and $Q$ anti-commute}~.
    \end{cases}
\end{align}
By using the commutation function and by taking the real projection of $\trans{P}{Q}$, we get 
\begin{align}\label{eq:trans_PQ_exact}
    \trans{P}{Q} = & \underbrace{- i h_{PQ} (\chi_{P,Q} -1)}_{ \trans{P}{Q}^{\rm coh}}
            + \underbrace{\sum_{j,S}   \left(\ell_{j,S} \ell_{j,PQS}^* \right) \left(\chi_{Q,S} - \frac{\chi_{P,Q}+1}{2}  \right)}_{\trans{P}{Q}^{\rm decoh}} ~,
\end{align}
{where we explicitly labeled the coherent and decoherent transition amplitudes.}
Let's break the transitions into two cases.\\

{\bf Case 1: $PQ=QP$}\\
 
First, if $PQ=QP$, then $\chi_{P,Q} =1$ and we get
 \begin{align}\label{eq:trans_PQ_commuting_exact}
    \trans{P}{Q} &=  \sum_{j,S}   \left(\ell_{j,S} \ell_{j,PQS}^* \right) \left(\chi_{Q,S} - 1  \right), \notag \\
                 &= -2\sum_{\substack{S \\ \chi_{Q,S} =-1 \\ \aoe{S+PQS}\leq k}}\sum_{j} \left(\ell_{j,S} \ell_{j,PQS}^* \right)~.
 \end{align}
In the special case where $P=Q$, we get
\begin{align}\label{eq:deriv_fidP}
    \trans{P}{P} = \frac{d}{dx}f_P(x)\Big |_0  = - 2 \sum_{\substack{S \\ \chi_{P,S}=-1 \\ \aoe{S} \leq k}}\sum_{j} |\ell_{j,S}|^2 = -\delta f_P^{\rm decoh}~.
\end{align}
Notice that the sum in one of the expressions above has 2 constraints, namely $\chi_{P,S}=-1$ and $\aoe{S} \leq k$. With these, the number of terms in the sum of \cref{eq:deriv_fidP} scales proportionally to the weight $w(P)$ of $P$, and scales exponentially in $k$ and in the level of connectivity of the architecture's graph $G$.

In \cref{eq:trans_PQ_commuting_exact} the transition amplitude is still scaling as the squared magnitude of the Lindbladian terms. {Applying the} Cauchy-Schwarz inequality {to the transition amplitude}, we get :
 \begin{align}
    |\trans{P}{Q}|
                 & = 2\left| \sum_{\substack{S \\ \chi_{Q,S} =-1\\
                 \aoe{S+PQS}\leq k}}\sum_{j} \left(\ell_{j,S} \ell_{j,PQS}^* \right) \right|~,\notag \\
                  & \leq 2 \sum_{\substack{S \\ \chi_{Q,S} =-1\\ \aoe{S+PQS}\leq k}}\sqrt{\sum_{j}|\ell_{j, S}|^2}\sqrt{\sum_j |\ell_{j, PQS}|^2}
\label{trans-amp-PQ}
 \end{align}
Because $PQ=QP$, we find that $PQS$ anti-commutes with $Q$ just like $S$, meaning that they belong to the same set of operators. This means that if $S$ is summed over an indexed set $\{S | \chi_{Q,S} =-1, \aoe{S+PQS}\leq k\} =: \{s_1,s_2, \cdots\}$, $PQS$ is simultaneously summed over a permutation 
 $\tau$ of that set:
\begin{align}
    |\trans{P}{Q}|
                  & \leq 2 \sum_{i}\sqrt{\sum_{j}|\ell_{j, s_i}|^2}\sqrt{\sum_j |\ell_{j, s_{\tau(i)}}|^2}~.
\label{trans-amp-PQ}
 \end{align}
It follows from majorization inequalities that the sum is maximized for the trivial permutation \cite{marshall11}:
\begin{align}
    |\trans{P}{Q}| &\leq 2 \sum_{\substack{S \\ \chi_{Q,S} =-1\\ \aoe{S+PQS}\leq k}} \sum_{j}|\ell_{j, S}|^2 = \delta f_Q^{\rm decoh}
\end{align}

The number of terms in the sum over $S$ is constrained since
\begin{enumerate}
    \item $S$ has to anti-commute with the transition endpoint $Q$ (i.e. $SQ=-QS$);
    \item  $S+PQS$ has to be geometrically $k$-local (i.e. $\aoe{S+PQS}\leq k$);
\end{enumerate}
Since $PQ \neq I$, the second condition implies that non-zero transitions must obey $\aoe{PQ} \leq k$; in simpler terms, $Q$ must differ from $P$ by a $k$-local operator. To get a better picture of non-zero transitions, imagine $P$ as a product of different ``Pauli islands'', defined as follows:
\begin{defn}[Pauli Islands]
A Pauli $P$ is said to be a $k$-island if any tensor product partitioning $P = Q \otimes R$ obeys 
\begin{align}
    \aoe{P|G}  < \aoe{Q|G} + \aoe{R|G}+k-1~.
\end{align}
\end{defn}
Any Pauli $P$ can be partitioned into a product of islands, and this product is unique.  For instance, for $k=2$ and a chain topology, $P= (X_2X_3X_4) \otimes (Z_{6}Y_{7})$ is a product of two islands. 
\begin{rules}[Transition rules for $PQ=QP$]\label{rule_commute}
The allowed transitions must obey the following rules:
\begin{enumerate}
    \item From the onset, $PQ=QP$ and $Q \neq P$.
    \item An allowed transition is the creation of a new geometrically $k$-local island. For instance with our $P= (X_2X_3X_4) \otimes (Z_{6}Y_{7})$ example, we could have an endpoint of the form
    $Q=(X_2X_3X_4) \otimes (Z_{6}Y_{7}) \otimes (Y_{75})$.
    \item An allowed transition can be the geometrically $k$-local modification of a single island. Still with our $P= (X_2X_3X_4) \otimes (Z_{6}Y_{7})$ example, we could have an endpoint of the form 
    $Q= (Y_1X_2X_3X_4) \otimes (Z_{6}Y_{7})$ or $Q= (X_2X_3X_4) \otimes (X_{6}X_{7})$.
    \item The annihilation of an island is forbidden! This is because of the first constraint on $S$, which states that it must anti-commute with $Q$, and enforces $S$ to connect to the endpoint $Q$.
    \item Through the second rule, islands that are less than $2k$ edges apart can be merged via island modifications. Still with our $P= (X_2X_3X_4) \otimes (Z_{6}Y_{7})$ example, we could have a single island endpoint: 
    $Q= (X_2X_3X_4Y_5Z_{6}Y_{7})$.
\end{enumerate}
\end{rules}

 By looking carefully at the above rules, notice that for a non-zero double transition $\trans{P}{Q}\trans{Q}{P}$ to start and end at $P$ and pass by
 $Q$ such that $QP=PQ$, it has to be a geometrically $k$-local modification of a single island (see \cref{rule_commute}, item 3). The reason for this is that although island creations have non-zero amplitudes, annihilation transitions are prohibited. Only a few triple transitions $\trans{P}{Q}\trans{Q}{R}\trans{R}{P}$ can start and end with $P$ and involve an island creation: 1) start with an island creation that is within $2k$ edges, 2) merge the created island to one of $P$'s islands in the second transition, 3) return to $P$ through a single island modification. 

With this argument in mind, let's bound the total amplitude of double transitions $\trans{P}{Q}\trans{Q}{P}$ that start and end at $P$ and pass by
 $Q$ such that $QP=PQ$:
 \begin{align}
    \sum_{\substack{Q, ~\chi_{Q,P}=1, \\ \text{geo. $k$-local}\\\text{mod. of 1 isl.} }} |\trans{P}{Q}||\trans{Q}{P}| \leq 4 \sum_{\substack{Q, ~\chi_{Q,P}=1, \\ \text{geo. $k$-local}\\\text{mod. of 1 isl.} }} \sum_{\substack{S \\ \chi_{Q,S} =-1\\ \aoe{S+PQS}\leq k}}  \sum_{\substack{S' \\ \chi_{P,S'} =-1\\ \aoe{S'+PQS'}\leq k}}\sum_{i,j} |\ell_{i, S}|^2 |\ell_{j, S'}|^2~.
\end{align}

Notice that the constrained double sum over $Q$ and $S$($S'$) goes over some $k$-local areas connected to $P$ and over some $k$-local jumps in those regions. Therefore, up to a multiplicity constant that grows exponentially in $k$, we get {
\begin{align}
    \sum_{\substack{Q, ~\chi_{Q,P}=1, \\ \text{geo. $k$-local}\\\text{mod. of 1 isl.} }} |\trans{P}{Q}||\trans{Q}{P}| &\leq O({\rm poly}(e^k)) \left[ |\trans{P}{P}| \max_{S \in \set S_k(P)} \left( \sum_{j} |\ell_{j, S}|^2\right) \right]~\notag \\
    & =  O({\rm poly}(e^k)) \left[ \delta f_P^{\rm decoh.} \max_{S \in \set S_k(P)} \left( e_S^{\rm decoh.}\right) \right]
\end{align}
}
which, unless the noise is highly non-local (i.e. $k$ is large), is much smaller than $|\trans{P}{P}|$. Triple transitions can be similarly neglected compared to $\trans{P}{P}$ on the first-order approximation.

 {\bf Case 2:$PQ=-QP$}
 \\
 
Finally, if $PQ=-QP$, we get
 \begin{align}
     \trans{P}{Q} = & 2 i h_{PQ}
            + \sum_{j,S}    \left(\ell_{j,S} \ell_{j,PQS}^* \right) \chi_{Q,S}
   ~.
 \end{align}
From there we can deduce some transition rules:
\begin{rules}[Transition rules for $PQ=-QP$]\label{rule_anticommute}
The allowed transitions must obey the following rules:
     \begin{enumerate}
         \item Since $P$ and $Q$ must anti-commute, island creation and annihilation are both forbidden.
        \item An allowed transition can {only} be the geometrically $k$-local modification of a single island.
     \end{enumerate}
 \end{rules}

Using the fact that $h_{PQ}=-h_{QP}$ and $\chi_{PQ,P} =-1$, we get:
\begin{align}
     \sum_{\substack{Q, ~\chi_{Q,P}=-1, \\ \text{geo. $k$-local}\\\text{mod. of 1 isl.} }} \trans{P}{Q}\trans{Q}{P} =&  - 4\sum_{\substack{S \\ ~\chi_{S,P}=-1}} |h_{S}|^2 + T 
\end{align}

where $T$ is a term with a magnitude bounded as
\begin{align}
     |T| = & \Bigg | \sum_{\substack{Q, ~\chi_{Q,P}=-1, \\ \text{geo. $k$-local}\\\text{mod. of 1 isl.} }} \sum_{\substack{S, S' \\  \aoe{S+PQS}\leq k \\\aoe{S'+PQS'} \leq k}}  \sum_{i,j}  \left(\ell_{i,S} \ell_{i,PQS}^* \right)   \left(\ell_{j,S'} \ell_{j,PQS'}^* \right) \chi_{Q,S}\chi_{P,S'} \notag \\
     & +2 \sum_{\substack{Q, ~\chi_{Q,P}=-1, \\ \text{geo. $k$-local}\\\text{mod. of 1 isl.} }}  i h_{PQ} \sum_{\substack{S \\  \aoe{S+PQS}\leq k }}  \sum_{j}  \left(\ell_{j,S} \ell_{j,PQS}^* \right)  \chi_{Q,S}\chi_{P,S'}\Bigg | \notag \\
     \leq & \sum_{\substack{Q, ~\chi_{Q,P}=-1, \\ \text{geo. $k$-local}\\\text{mod. of 1 isl.} }} \left(2 \left|h_{PQ}  \right|+\sum_{\substack{S\\  \aoe{S+PQS}\leq k}}  \sum_{j} |\ell_{j,S}|^2  \right)  \left( \sum_{\substack{S \\  \aoe{S+PQS}\leq k }}  \sum_{j}  |\ell_{j,S}|^2 \right)~,\notag \\
     \leq & \max_{\substack{Q, ~\chi_{Q,P}=-1, \\ \text{geo. $k$-local}\\\text{mod. of 1 isl.} }} \left(2 \left|h_{PQ}  \right|+\sum_{\substack{S\\  \aoe{S+PQS}\leq k}}  \sum_{j} |\ell_{j,S}|^2  \right)  \sum_{\substack{Q, ~\chi_{Q,P}=-1, \\ \text{geo. $k$-local}\\\text{mod. of 1 isl.} }} \sum_{\substack{S \\  \aoe{S+PQS}\leq k }}  \sum_{j}  |\ell_{j,S}|^2 \label{eq:bound_complicated}
\end{align}

where the second line was obtained by using
\begin{align}
    \left| \sum_{j,S}   \left(\ell_{j,S} \ell_{j,PQS}^* \right) \chi_{Q,S} \right| \leq  \sum_{\substack{S \\ \aoe{S+PQS}\leq k}} \sum_{j}|\ell_{j, S}|^2~.
\end{align}
 The second factor in \cref{eq:bound_complicated} scales as the average single transition $\trans{Q}{Q}$ over all Paulis $Q$ with the same support as $P$:
\begin{align}
     \sum_{\substack{Q, ~\chi_{Q,P}=-1, \\ \text{geo. $k$-local}\\\text{mod. of 1 isl.} }} \sum_{\substack{S \\  \aoe{S+PQS}\leq k }}  \sum_{j}  |\ell_{j,S}|^2 = O({\rm poly}(e^k)) \times e_{\set S_k(P)}^{\rm decoh}~.
\end{align}
Using \cref{eq:deriv_fidP}, this means that $|T|$ can be bounded by 
\begin{align}
    |T| \leq&   O({\rm poly}(e^k)) \times \left(2\max_{\substack{S, ~\chi_{S,P}=-1 }}  \left|h_{S}  \right|\right)  \times  e_{\set S_k(P)}^{\rm decoh} \notag \\
& +  O({\rm poly}(e^k)) \times \left(\max_{\substack{S \in \set S_k(P)}}  e_S^{\rm decoh.} \right) \times  e_{\set S_k(P)}^{\rm decoh}~.
\end{align}

\end{proof}

\subsection{Derivation of the Error Probabilities of a repeated channel}

\label{sec:appendix B}
Error processes are not necessarily stochastic, meaning that it doesn't always make sense to discuss error probabilities. However, we can always project the error channels unto a stochastic one, and then consider the resulting error probability distribution. As such, we define the effective Pauli error probabilities as the resulting Pauli error probability of a channel once it is projected onto its Pauli stochastic component.

When it comes to Pauli stochastic channels, there is a duality between Pauli fidelities and Pauli error probabilities. The two are in fact related by a linear operation known as the Walsh-Hadamard transform. That is, if we consider a vector of fidelities, $\mathbf{f}$ (such as $\mathbf{f}= (f_I, f_X,f_Y,f_Z)$), we can obtain the vector of error probabilities $\mathbf{e}$  (such as $\mathbf{e} = (e_I, e_X,e_Y,e_Z)$) via:
\begin{align}\label{wh_transform}
    \mathbf{e} = W \mathbf{f},
\end{align}
where $W$ is the Walsh-Hadamard matrix. The entry $W_{ij}$ is $1$ if the $j$th Pauli in the domain vector commutes with the $i$th Pauli in the image. $W_{ij}$ is $-1$ if the $j$th Pauli in the domain vector anti-commutes with the $i$th Pauli in the image.

Therefore, we can get error probabilities from fidelities by using the Walsh-Hadamard transform on the fidelities given by \cref{lem:fid_quad}. Let's approximate the elements of the fidelity vector 
\begin{align}
    f_P(x) \simeq 1-\delta f_P^{\rm coh.} x^2 -\delta f_P^{\rm decoh} x~.
\end{align}
If we apply the Walsh-Hadamard transform $W$ on the vector composed of these approximated elements, we get an approximate error vector $\mathbf{e}$ with elements
\begin{align}
   {e}_P(x)  \simeq x^2 |h_{P}|^2 +x  \sum_j |\ell_{j,P}|^2~.
\end{align}
More mathematical details regarding the transformation form fidelities to effective error rates is contained in \cite{CER_2023}.

\newpage

\section{IBM Quantum Computing Hardware Architectures}
\label{appendix:IBM-hardware}

The calculations for this project were run on three different IBM hardware platforms (\texttt{ibmq\_manila}, \texttt{ibmq\_guadalupe}, and  \texttt{ibmq\_montreal}).  {On the 5 qubit \texttt{ibmq\_manila} platform, qubits q0, q1, q2, q3, and q4 are used with q0 as the ancilla and q1-q2 as the two-qubit entangling gate.}  The \texttt{ibmq\_guadalupe} processor is a 16 qubit platform.  On this processor the qubits used are q6, q7, q10, q12, and q13 with q10 as the ancilla and q6-q7 as the two-qubit entangling gate.  {For the computation done on the 27 qubit  \texttt{ibmq\_montreal} platform, qubits q18, q21, q23, q24, and q25 are used with q23 as the ancilla and q18-q21 as the two-qubit entangling gate.} 
\cref{fig:IBM hardware} graphically shows these qubit topologies for each platform.  We also ran computations on the Keysight TrueQ simulator \cite{trueq} for the analysis discussed in the supplemental material in \cref{sec:fitting}.

\begin{figure*}[!htpb]
    \centering
    \includegraphics[width=5.0in,height=3.0in]{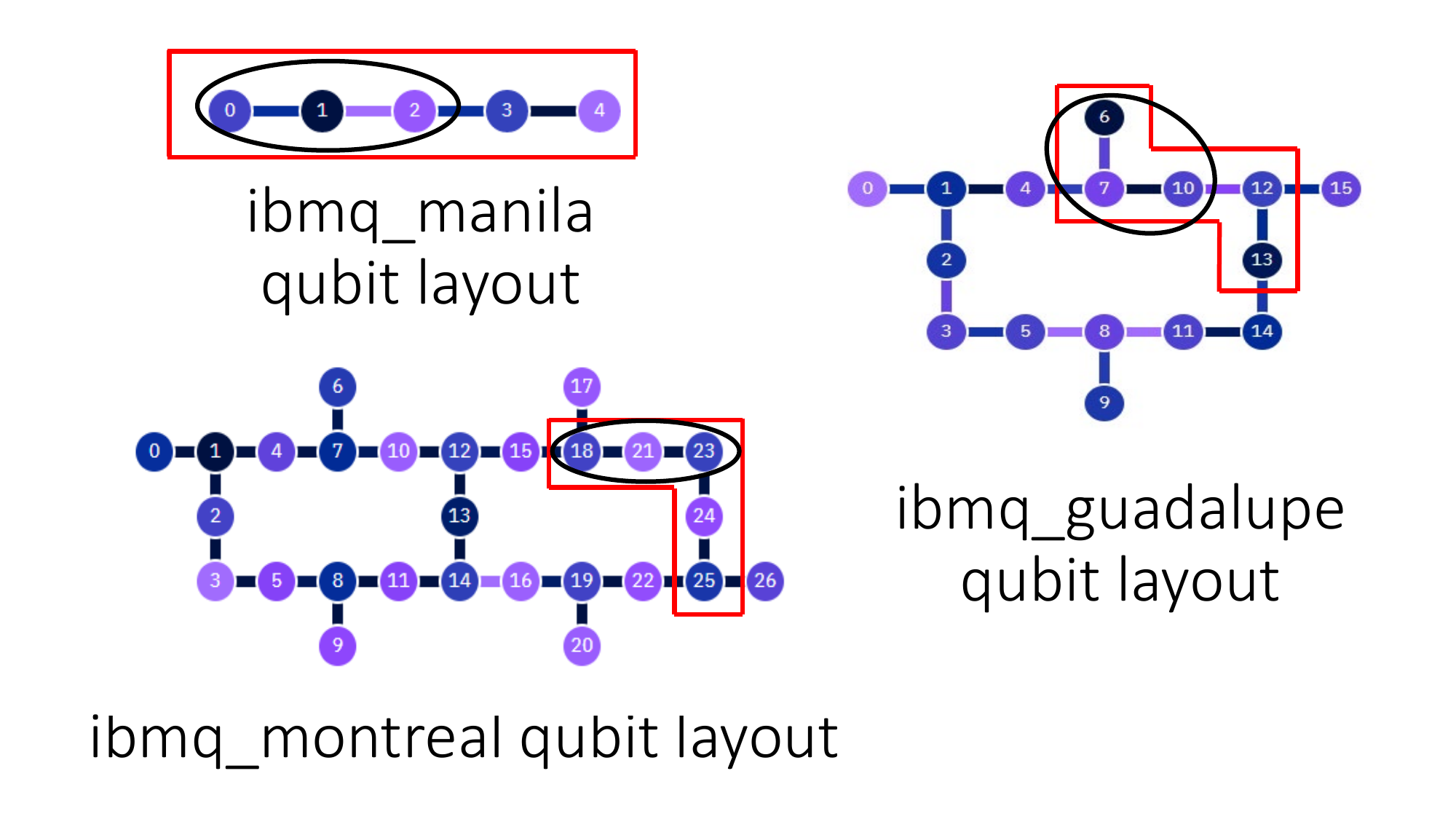}
\caption{IBM quantum hardware platforms and specific qubits used on each to run the 4 qubit spin-spin correlation function circuit.  The black circles indicate the specific qubits used for the KNR two-qubit CNOT and ancilla effective dressed cycles and Pauli fidelity calculations}
\label{fig:IBM hardware}
\end{figure*}

The key metrics associated with each of the processors that help characterize the processor performance are
\begin{itemize}
    \item Quantum volume (QV) This value measures the performance of gate-based quantum computers, regardless of their underlying technology.
    \item Circuit Layer Operations per Second (CLOPS) is a measure of how many layers of a QV circuit a quantum processing unit (QPU) can execute per unit of time. The CLOPS is calculated using three key attributes to measure the performance of near-term quantum computers (quality, speed, and scale)~\cite{CLOPS}.
    \item The Falcon family of devices are medium-scale circuits.  They were deployed by IBM as a test environment for demonstrating performance and scalability improvements over previous generation processors.  Specifically the r4 is the first revision to deploy multiplexed readout. Previous designs required an independent signal pathway on the chip, as well as in the dilution refrigerator and control electronics for qubit state readout.
\end{itemize}

The \texttt{ibmq\_guadalupe} is a 16 qubit Falcon r4p system with a quantum volume of 64 and 2.4K CLOPS.  The computations on Guadalupe use qubits 6, 7, 10, 12, 13 with qubit 6 being used as the ancilla qubit.   The \texttt{ibmq\_montreal} is a 27 qubit Falcon r4 system with a quantum volume of 128 and 2.0K CLOPS.

\section{CER Protocol Measurements and Heatmap Methodology}
\label{appendix:CER-Heat}

This project measures the noise on the single qubit (q10 in \texttt{ibmq\_guadalupe}, q23 for \texttt{ibmq\_montreal} and q0 for \texttt{ibmq\_manila}).  
 Although the single qubit gates are set to the identity and should remain so despite the increase in noise from the additional CNOT folded hard cycles implemented through the extended CER mitigation protocol, these heatmaps show that what is measured is not what was expected.   

{
The increase in the measured noise on the single qubit is graphically seen in ~\cref{fig:guad_heatmap}-\cref{fig:manila_heatmap}).  These figures are heatmap error signatures showing the magnitude of error recorded on these qubits.  A dark color represents a relatively low value for the error; warmer colors represent a higher value.  To the right of each figure is a bar with a color gradation and numbers that set the scale for that heatmap representing the level of the infidelity being measured.   
}

{
Each of the three figures (~\cref{fig:guad_heatmap} - \cref{fig:manila_heatmap}) shows five sub-figures representing the single qubit error measurements for CNOT hard cycle repetitions 1, 3, 5, 7, and 9 for each of the three IBM hardware platforms.  The y-axis label of the heatmap shows the X, Y, and Z Pauli errors for the qubit of interest (q0 for \texttt{ibmq\_manilla}, q10 for \texttt{ibmq\_guadalupe} and q23 for \texttt{ibmq\_montreal)}.  
}

{
These heatmaps also clearly show that as the number of hard cycles is increased from 1 to 9, the magnitude of the single qubit Pauli errors (especially the Z error) grows by an order of magnitude on each of the three different hardware platforms.  This is a graphical signature that the CNOT folded cycles are contributing an increasing magnitude of error that is detected by the deviation of the single spectator qubit from what should have been an identity gate signature. 
}

\begin{figure}[h]
   \centering
   \begin{tabular}{@{}c@{\hspace{.5cm}}c@{}}
       \includegraphics[page=1,width=.45\textwidth]{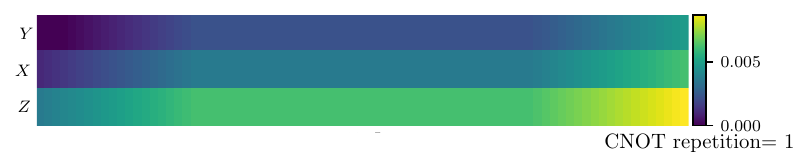} & 
       \includegraphics[page=2,width=.45\textwidth]{Guad_heatmap.pdf} \\[.5cm]
       \includegraphics[page=3,width=.45\textwidth]{Guad_heatmap.pdf} &
       \includegraphics[page=4,width=.45\textwidth]{Guad_heatmap.pdf} \\[.5cm]
       \includegraphics[page=5,width=.45\textwidth]{Guad_heatmap.pdf} \\
   \end{tabular}
 \caption{Heatmap of \texttt{ibmq\_guadalupe} processor. To the right of each figure is a bar with a color
gradation and numbers that set the scale for that heatmap representing the level of the
infidelity being measured}
 \label{fig:guad_heatmap}
\end{figure}

\begin{figure}[h]
   \centering
   \begin{tabular}{@{}c@{\hspace{.5cm}}c@{}}
       \includegraphics[page=1,width=.45\textwidth]{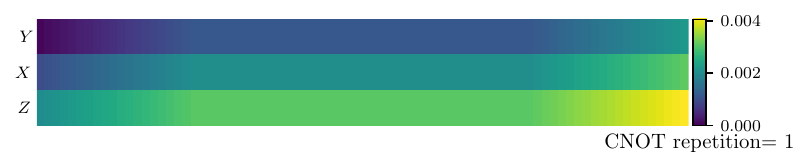} & 
       \includegraphics[page=2,width=.45\textwidth]{Montreal_heatmap.pdf} \\[.5cm]
       \includegraphics[page=3,width=.45\textwidth]{Montreal_heatmap.pdf} &
       \includegraphics[page=4,width=.45\textwidth]{Montreal_heatmap.pdf} \\[.5cm]
       \includegraphics[page=5,width=.45\textwidth]{Montreal_heatmap.pdf} \\
   \end{tabular}
 \caption{Heatmap of \texttt{ibmq\_montreal} processor. To the right of each figure is a bar with a color
gradation and numbers that set the scale for that heatmap representing the level of the
infidelity being measured}
 \label{fig:montreal_heatmap}
\end{figure}

\begin{figure}[h]
   \centering
   \begin{tabular}{@{}c@{\hspace{.5cm}}c@{}}
       \includegraphics[page=1,width=.45\textwidth]{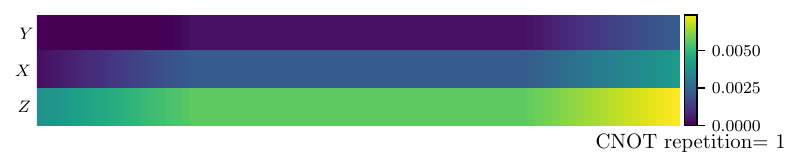} & 
       \includegraphics[page=2,width=.45\textwidth]{manilla_heatmap.pdf} \\[.5cm]
       \includegraphics[page=3,width=.45\textwidth]{manilla_heatmap.pdf} &
       \includegraphics[page=4,width=.45\textwidth]{manilla_heatmap.pdf} \\[.5cm]
       \includegraphics[page=5,width=.45\textwidth]{manilla_heatmap.pdf} \\
   \end{tabular}
 \caption{Heatmap of \texttt{ibmq\_manila} processor. To the right of each figure is a bar with a color
gradation and numbers that set the scale for that heatmap representing the level of the
infidelity being measured}
 \label{fig:manila_heatmap}
\end{figure}

\newpage

\end{document}